\documentclass[twocolumn,amsmath,amssymb,prl,aps,showkeys]{revtex4}
\usepackage{graphicx}
\usepackage{amsmath,amssymb}
\usepackage{mathrsfs}
\usepackage{color}
\usepackage{afterpage}
\usepackage[version=3]{mhchem}
\usepackage[normal]{subfigure}
\usepackage{natbib}
\usepackage{soul}

\begin{document}

\preprint{APS/123-QED}

\title{Self energy and excitonic effect in (un)doped TiO$_2$ anatase : A comparative study of hybrid DFT, GW and BSE to explore optical properties }
\author{Pooja Basera$^*$, Shikha Saini, Saswata Bhattacharya\footnote{Pooja.Basera@physics.iitd.ac.in [PB], saswata@physics.iitd.ac.in [SB]}} 
\affiliation{Dept. of Physics, Indian Institute of Technology Delhi, New Delhi 110016, India}




\date{\today}

\begin{abstract}
\noindent
TiO$_2$ anatase has its significant importance in energy and environmental research. However, the major drawback of this immensely popular semi-conductor is its large bandgap of 3.2 eV. Several non-metals have been doped experimentally for extending the TiO$_2$ photo-absorption to the visible region. Providing in-depth theoretical guidance to the experimentalists to understand the optical properties of the doped system is therefore extremely important. We report here using state-of-the-art hybrid density functional approach and many body perturbation theory (within the frame work of GW and BSE) the optical properties of p-type (S and Se doped) and n-type (N and C doped) TiO$_2$ anatase. The anisotropy present in non-metal doped TiO$_2$ plays a significant role in the optical spectra. The p-type dopants are optically active only for light polarized along xy direction, whereas the n-type dopants are optically active when light is polarized along xy and z direction in low energy region. We have found that, in all the doped systems optically allowed transitions are introduced well below 3 eV (i.e. visible spectra region). This helps to improve its opto-electronic and solar absorption properties. All the calculations are well validated with respect to the available experimental observation on pristine TiO$_2$ anatase. 
\end{abstract}

\pacs{Valid PACS appear here}
\maketitle


\section{Introduction}
Anatase TiO$_2$ has attracted worldwide research sensation because of its wide applications in the field of opto-electronic devices and solar cell conversion efficiency~\cite{chen2014electrospun,bai2014titanium,gratzel2005solar,o1991low}. High carrier mobility, chemical stability, low recombination rate and cheap manufacturing cost have made this material very popular for the applications in photo-active and opto-electronic devices~\cite{chen2014electrospun,bai2014titanium}. However, anatase TiO$_2$ has wide band gap (3.2 eV) that limits its response in the visible region. Hence, doping is often employed to extend the photo-activity of TiO$_2$ in the visible region (1.5 - 2.8 eV) (besides UV region)~\cite{ghosh2012ab}. Different types of dopants viz. transition metals~\cite{kim2004high, liu2013large, wang2014first, janisch2005transition}, noble metals~\cite{ribao2017tio2, ma2014noble}, charge-compensated and uncompensated co-doped elements~\cite{zhu2009band, wang2014band}  have been reported in past. Unfortunately, most of these metal dopants often act as recombination centre owing to their capacity to induce trapping levels inside the band gap. As a consequence of this, the latter reduces a significant amount of photo-excited carriers~\cite{pelaez2012review}. 

To avoid such additional trapping levels inside the band gap region, non-metal doping is suggested as an alternative strategy to improve visible-light absorption~\cite{irie2003nitrogen, irie2003carbon, li2005synthesis, asahi2001visible, chen2004photoelectron, catal_today_2013, chem_mat_2005, ohno2003photocatalytic}. Various non-mental dopants X, (X = nitrogen (N), carbon (C), sulphur (S) and selenium (Se)) have already been experimentally reported~\cite{varley2011mechanism, wang2009simple, liu2010sulfur, xie2018enhanced}. We have recently established the enhanced thermodynamic stability of charged substitutional defects (X$_\textrm{O}^q$) (as compared to interstitials and other kind of defects) of these non-metal dopants at O-atom site in TiO$_2$ anatase~\cite{PB}.  In this context, theoretical simulation of a detailed analysis of the influence of dopants on the optical absorption can contribute new insights of scientific and practical interest as well.

Until date, there are few theoretical studies where optical properties of pristine TiO$_2$ is addressed~\cite{chiodo2010self,zhu2014stability}. However, for doped TiO$_2$ there are not many reports to address the optical properties in the context of its effective application in optoelectronic devices. To study the optical properties of doped TiO$_2$, we need an advanced computational method, which is reliable enough to evaluate the precise band gap and excited states. Note that despite the huge success in predicting the ground state properties, density functional theory (DFT) with  local / semi-local functionals (viz. LDA~\cite{fuchs2005comparison} or GGA~\cite{perdew1996generalized}), is not sufficient to determine the excited state properties. To circumvent this problem DFT+U and/or hybrid functionals are sometimes effective in determining the structure and energetics of the excited states~\cite{hubbard1963electron}. However, in many instances these approaches are also insufficient for precisely describing the electronic energy levels~\cite{gerosa2017accuracy}. For instance, hybrid functionals often overestimate the TiO$_2$ band gap thus leading to an uncertainty in the position of the impurity levels~\cite{gerosa2017accuracy}. Recent many-body perturbation theory (MBPT) studies of the band structure and absorption spectrum of pristine TiO$_2$, have ensured that approaches beyond DFT are essential for a quantitative description of photoemission, inverse photoemission and light absorption experiments in this material~\cite{kang2010quasiparticle,chiodo2010self,zhu2014stability}. Therefore, we require an advanced computational method like many body perturbation theory (MBPT), which includes first order (GW)~\cite{chiodo2010self,jiang2010first,karsai2014f,onida2002electronic} and higher order Bethe-Salpeter Equation (BSE)~\cite{fuchs2008efficient, salpeter1951relativistic} Green's function approaches to calculate the excited state properties. 

Note that, we have already established that GGA+U is not a suitable functional for this system~\cite{PB}. In the present article, firstly we have adopted configuration coordinate (CC) approach using hybrid functional HSE06 to calculate the optical absorption and emission peaks in doped TiO$_2$. Next, we have discussed the need of many body perturbation theory approaches (GW and BSE). In this section, we have first validated the proper choice of starting DFT functional. Note that, here the calculations are restricted to single shot GW. It is well known that single shot GW calculations are highly dependent on the starting point. Thus, it is important to benchmark thoroughly the starting point of this calculation. Finally, we have studied the optical properties of pristine and doped TiO$_2$. 

\section{Methodology}
DFT calculations are performed with PAW pseudopotential method~\cite{blochl1994projector} as implemented in Vienna {\em ab initio} simulation package (VASP)~\cite{kresse1996efficient}. The size of the tetragonal TiO$_2$ unit cell (space group: I41/amd) is kept on increasing until the single defect state becomes fully localized inside the supercell. All the structures (i.e. atomic positions) are relaxed upto 0.001 eV/$\textrm{\AA}$ force tolerance using conjugate gradient minimization algorithm with 4$\times$4$\times$2 k-mesh. For electronic structure energy calculations, the brillouin zone is sampled with a 8$\times$8$\times$4 Monkhorst-Pack~\cite{monkhorst1976special} k-mesh. The energy tolerance is kept at 0.01 meV throughout, whereas the plane wave energy cut-off is set to 600 eV. The excited state calculations are performed with GW approach, which involves the screened coulombic interaction and improves the Hartree Fock approximation. 
For quasi-particle calculations, one should be very careful while choosing the convergence criteria for excited states. The calculations are performed on a grid of 50 frequency points, and this no. is thoroughly benchmarked. A more detailed description of the method can be found in Refs.~\cite{hedin1965new, hybertsen1985ms}.
\begin{figure}
	\includegraphics[width=0.5\textwidth]{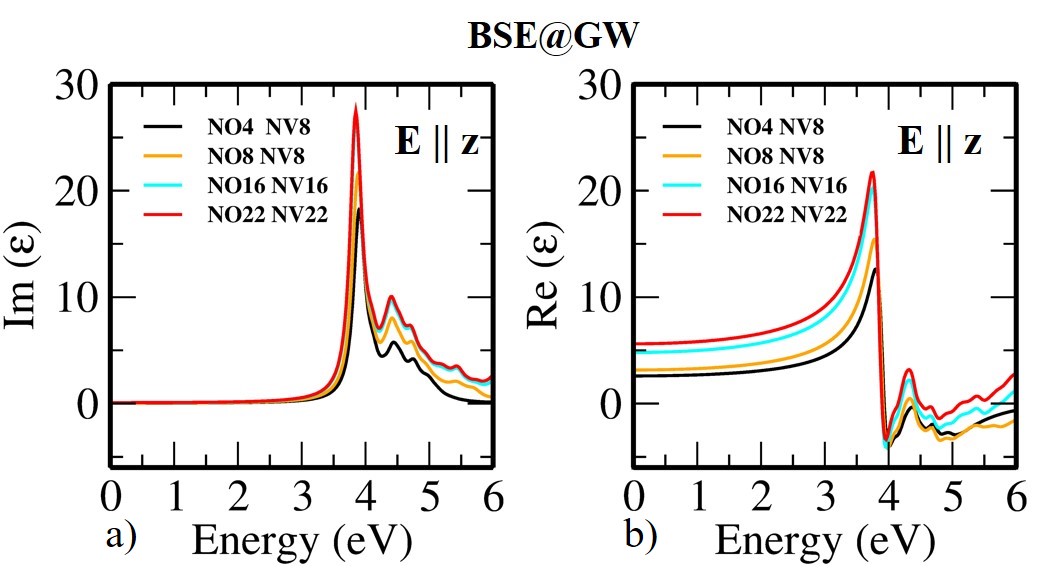} 
	\caption{(a) Imaginary and (b) real part of the dielectric function as calculated for a 6$\times$6$\times$6  k-mesh, by including the different number of occupied (NO) and unoccupied bands (NV). For NO=NV=22, the static value of real part of dielectric function is 5.61, which has a nice match with the experimental value i.e. 5.62~\cite{wemple1977optical}.}
    \label{fig1}
\end{figure}
The second order green's function technique i.e., BSE is also employed on top of GW to include the excitonic effects. 
For BSE calculations the real and imaginary part of the dielectric function are carefully benchmarked with respect to different numbers of occupied (NO) and unoccupied bands (NV). In Fig~\ref{fig1} the imaginary and real part of the dielectric function for different numbers of NO and NV [viz. NO = 4, NV = 8; NO = 8, NV = 8; NO = 16, NV = 16; NO = 22, NV = 22 etc.] are calculated for polarisation along the z direction of pristine TiO$_2$. Fig~\ref{fig1}a shows slight change in imaginary part of the dielectric function (only intensity of the peak), whereas it is clear from Fig~\ref{fig1}b that there is a significant change in  static value of real part of the dielectric function. In the case of NO = 22 and NV = 22,  the static value of  real part of dielectric constant (5.61) is consistent with the experimental value (5.62)~\cite{wemple1977optical}. Therefore, for rest of the calculations we have set NO and NV at 22. 
A more detailed discussion on BSE calculations is given in Ref.~\cite{onida2002electronic}.
	
\section{Results}
\subsection{Configuration co-ordinate (CC) diagram : Hybrid functional approach}
\begin{figure}
	\includegraphics[width=0.4\textwidth]{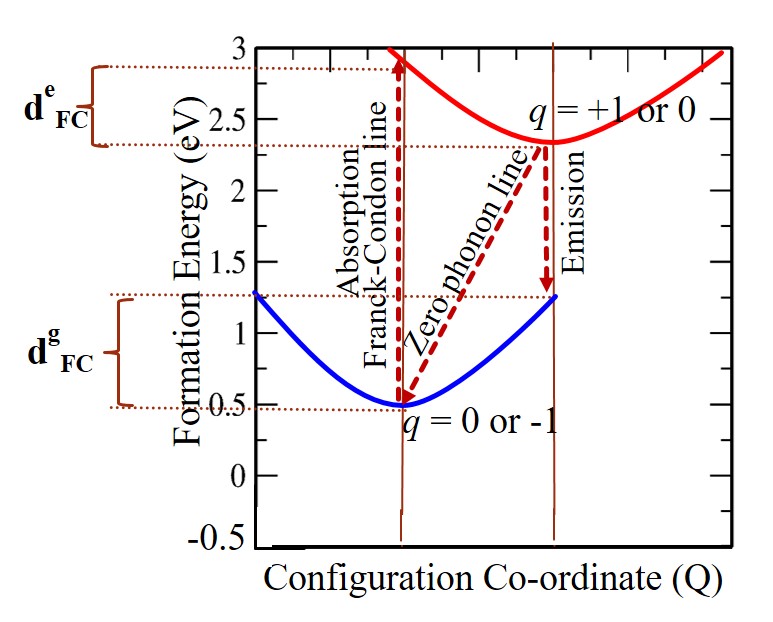} 
	\caption{Schematic picture of the defect geometry associated with the change of its electronic state (charge of the defect) is described by a configuration coordinate diagram. It shows the ground and excited state of the  charged (neutral) defect as a function of an effective coordinate. The vertical arrow indicate the Franck Condon line whereas nonvertical transition corresponds to thermal energy of the transition i.e zero phonon line which is defined as the energy difference between the ground state and the excited state in their equilibrium configurations.}
	\label{fig2}
\end{figure}

The variation of the electronic potential surface along the various geometries (with specific charge states) is shown by a schematic configuration co-ordinate (CC) diagram (see Fig~\ref{fig2}). The difference between the Franck-Condon line (vertical transition) and the zero phonon line (non-vertical transition) is defined as the Franck-Condon shift d$_{\textrm{FC}}$  or relaxation energy of the excited state~\cite{karsai2014f,lyons2014effects}. This is an upper bound to the contribution of electron-phonon to the absorption energy. The Franck-Condon shift for the ground (excited) state is represented by d$_{\textrm{FC}}^{g}$ (d$_{\textrm{FC}}^{e}$) as shown in Fig~\ref{fig2}. The zero-phonon line (ZPL) corresponds to the excitation from the minimum of the ground state to the minimum of the excited-state energy (equilibrium position of ground and excited state). Hence, it corresponds to the lowest possible absorption energy (see Table~\ref{Table}).
\begin{table}[htbp]
	\caption{Franck Condon shift in ground state d$_{\textrm{FC}}^{g}$ and in excited state   d$_{\textrm{FC}}^{e}$  of non-metal doped TiO$_2$ and their ZPL.} 
	\begin{center}
		\begin{tabular}[c]{|c|r|c|r|c|r|} \hline
			Dopants & d$_{\textrm{FC}}^{g}$ (eV) & d$_{\textrm{FC}}^{e}$ (eV) & ZPL (eV)\\ \hline
			S$_\textrm{O}$   & 0.49  & 0.58 & 1.74 \\ \hline
			Se$_\textrm{O}$  & 0.48  & 0.56 & 1.36  \\ \hline
			N$_\textrm{O}$  & 0.59  & 0.60 & 1.54  \\ \hline
			C$_\textrm{O}$  & 0.51  & 0.58 & 0.98  \\ \hline
		\end{tabular}
		\label{Table}
	\end{center}
\end{table}

\begin{figure}
	\includegraphics[width=0.5\textwidth]{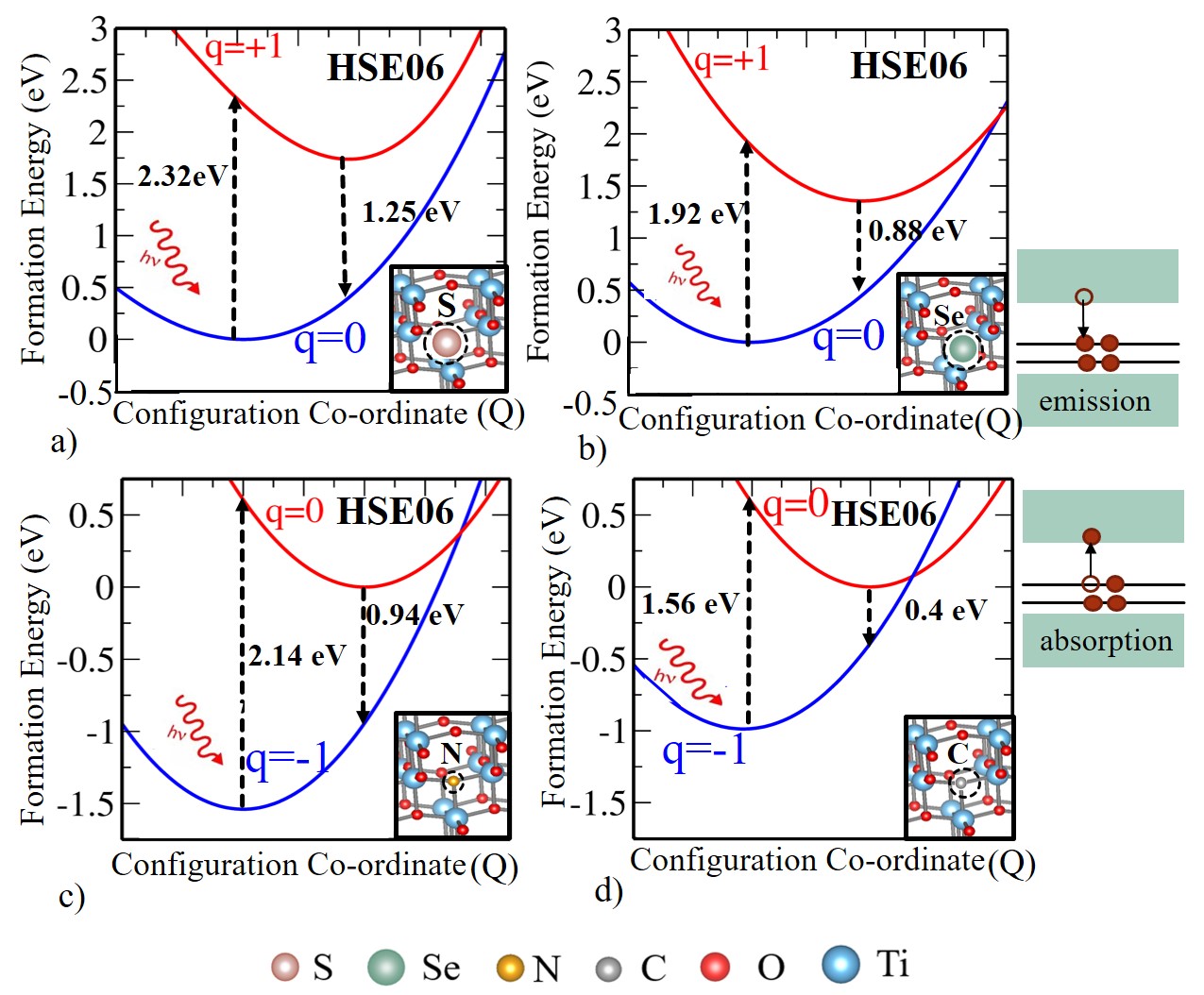} 
	\caption{Configuration coordinate diagrams for X$_\textrm{O}$ (X= S, Se, N and C) in anatase TiO$_2$. The formation energies in two different charge state (\textit{q}=0 and \textit{q}=+1) are plotted as a function of the displacement of atoms.}
	\label{fig3}
\end{figure}

The optical transitions associated with the process, from charge (\textit{q} = 0 to \textit{q} = +1) or (\textit{q} = -1 to 0) i.e., X$_\textrm{O}^0$ + h$\nu$ $\rightarrow$ X$_\textrm{O}^+$ + e$^{-}$ or X$_\textrm{O}^-$ + h$\nu$ $\rightarrow$ X$_\textrm{O}^0$ + e$^{-}$  are shown in Fig~\ref{fig2} (schematic) and~\ref{fig3}. To do this, we have taken three different stable geometry configurations of X$_\textrm{O}$ at three respective charge states viz. \textit{q} as 0, +1 and +2 for donor type of dopant (e.g. S, Se), whereas \textit{q} as 0, -1 and -2 for acceptor type of dopant (e.g. N, C). Therefore, S$_\textrm{O}$ and Se$_\textrm{O}$ are donor type of dopants, whereas N$_\textrm{O}$ and C$_\textrm{O}$ are acceptor type of dopants. The formation energies of these  configurations are calculated using HSE06 xc-functional (with $\alpha = 0.22$, see SI of Ref.~\cite{PB} for validation of $\alpha$). The formation energy data points of the configuration coordinates are fitted using interpolation method to obtain the smooth curve of the potential energy surface. The scientific reason for this interpolation fitting is that bond is assumed to behave like a harmonic spring upto first order, as E $\sim$ Q$^2$. The possible transitions are illustrated in Fig~\ref{fig2}. Once the optical transitions are known, the corresponding optical spectra can be computed. The details of this state-of-the-art theoretical spectroscopy techniques are explained by Rinke et al.~\cite{rinke2012first}. The peak positions are determined using the Franck-Condon principle, assuming that atoms surrounding the defect do not have time to relax during the transition from charge state $q$ to charge state \textit{q$'$}(= \textit{q} $\pm$ 1).  Franck-Condon principle states that, when molecule undergoes an electronic transition such as ionization, there is no significant change in nuclei configuration. This is obvious as nucleus being heavier than electron, the electronic transitions are faster. Thus, the difference in formation energies in charge states \textit{q} and \textit{q$'$} gives the absorption and emission energy. The absorption energy is defined as:
E(absorption) = ZPL + d$_{\textrm{FC}}^{e}$ (for values see Table~\ref{Table}). The difference in the value of absorption and emission peaks arises due to difference in geometry relaxation for both the charge states \textit{q} and \textit{q$'$}. The amount of energy loss during relaxation to the new equilibrium position is the Franck-Condon shift, d$_{\textrm{FC}}$.  This leads to significant stokes shifts and strong vibrational broadening of luminescence peaks. Thus, the total energy difference between absorption and emission is called the stokes shift~\cite{alkauskas2016tutorial}. Therefore, the stokes shift is simply the sum of Franck-Condon shifts in the ground and excited state:\\
E(absorption)-E(emission) = d$_{\textrm{FC}}^{g}$ + d$_{\textrm{FC}}^{e}$\\ The Optical transition levels are usually determined in photoluminescence (PL) or cathodoluminescence (CL) experiments~\cite{karsai2014f}.

Knowing the basic terms used in the CC diagram, we proceed towards the detailed study of optical transition for various non-metal dopants (X = S, Se, N and C) at O-atom site.
The optical transitions associated with the process S$_\textbf{O}^0$+h$\nu$ $\rightarrow$ S$_\textbf{O}^+$+e$^{-}$ corresponds to absorption of photon and emission of electron (see Fig~\ref{fig3}a). Note that we have shown optical transition for (0/+) state instead of ($-$1/0). The cognition for aforementioned statement is already given in our earlier work, where we have proved on the basis of formation energy that sulphur prefers to be in cationic state rather than anionic state~\cite{PB}. We have found that S$_\textbf{O}$ leads to sub-band gap optical absorption transitions centered at 2.32 eV (534 nm), which lies within the visible region. However, the emission peak lies at 1.25 eV (991.8 nm). The  difference between absorption and emission peaks arises due to different local lattice relaxation for the geometries S$_\textbf{O}^0$ and S$_\textbf{O}^+$. The absorption results are in good agreement with the experimental observations~\cite{li2007supercritical}.

The optical transitions for Se$_\textbf{O}^0$+h$\nu$ $\rightarrow$ Se$_\textbf{O}^+$+e$^{-}$  is shown in Fig~\ref{fig3}b. Like sulphur, selenium also favors the cationic state \textit{q} $=$ + 1, + 2~\cite{PB}. Thus, we have carried out calculations for the optical transition corresponding to (0/+1) charge state. We find Se$_\textbf{O}$ leads to sub-bandgap optical absorption peak at 1.92 eV (645 nm), which also lies within the visible region. However, emission peak lies at 0.88 eV (1408 nm). The different local lattice relaxation of the geometries Se$_\textbf{O}^0$ and Se$_\textbf{O}^+$ is responsible for the huge difference between absorption and emission peaks ~\cite{xie2018enhanced}. 

The optical transitions of the N$_\textrm{O}^-$ defect is shown in Fig~\ref{fig3}c. Note that, unlike S and Se, here we have shown optical transition for ($-$1/0) charge state due to the fact that N$_\textrm{O}^-$ (with one electron trapped) is the most stable configuration possible for various N related defects~\cite{PB}. 
We have obtained that N$_\textrm{O}$ leads to sub-bandgap optical transitions, where absorption is centered at 2.14 eV (579nm) and the emission peak lies at 0.91 eV (1362 eV) as shown in Fig~\ref{fig3}c. These results are in good agreement with the experimental observations in which sizeable absorption occurs at (400-600 nm)~\cite{cheung2008n}. 

Finally we evaluate the optical properties of the C$_\textrm{O}^-$ defects. Like N, C also prefers the anionic state. Since, carbon has two electron less than oxygen. Thus, it has a tendency to accept an electron and form  C$_\textrm{O}^-$, C$_\textrm{O}^{-2}$~\cite{PB}. The optical transitions associated with the process C$_\textrm{O}^-$+h$\nu$ $\rightarrow$ C$_\textrm{O}^0$+e$^{-}$ is shown in Fig~\ref{fig3}d. We have found that C$_\textrm{O}$ leads to absorption peak at 1.56 eV (794nm), which lies in the visible and near-infrared region~\cite{lin2011carbon}.

Note that, the CC diagram is only limited to obtain the optical absorption and emission peaks of the defected system (but is not applicable for pristine TiO$_2$). Moreover, here we have explicitly discussed the experimental range (400-800 nm), in which the absorption is most likely to occur. However, the exact location of optical transition absorption peaks (which includes the effect of electron-hole) has not been taken into account. The dopants (X) significantly perturb the system of TiO$_2$ anatase and therefore, results in complicated spectrum of quasi-local modes. The broadening of the spectra has not been considered in the CC diagram. Hence, it is not able to capture the fine spectra, which is observed in experiments. Later, we will show the inclusion of electron-hole (excitonic) effects transpire the  fine spectra close to experiments. Also, the detailed optical properties can be determined only if one knows the real and imaginary part of the dielectric function. Therefore, for the accurate description of real and imaginary part of dielectric function, we need an advanced computational approach based on many body perturbation theory i.e., GW and BSE. 
\subsection{GW and BSE approach}
\begin{figure}	
\includegraphics[width=0.5\textwidth]{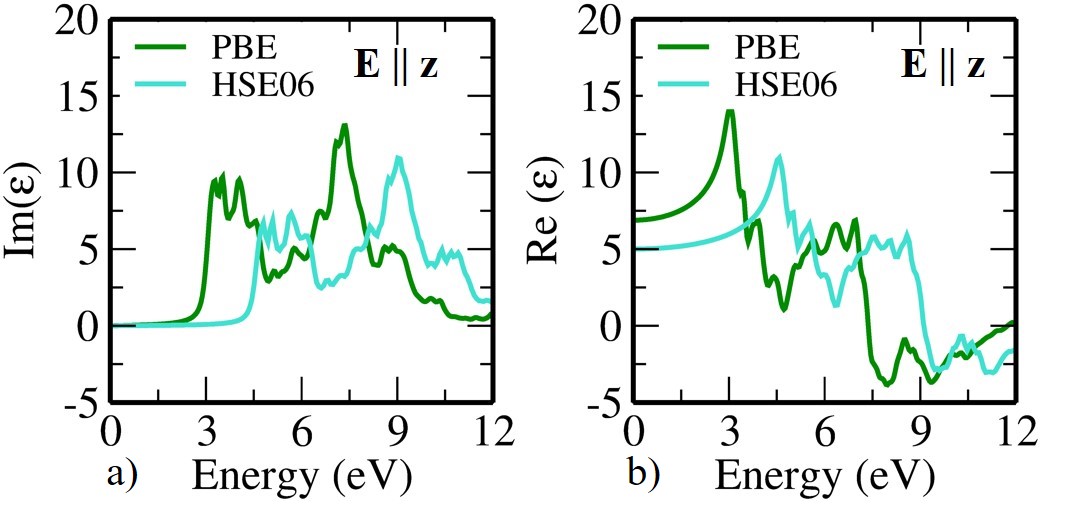} 
\caption{ (a) Imaginary and (b) real part of the dielectric function using PBE and HSE06 functional along z direction. The experimental value for the static real part of the dielectric function is 5.62~\cite{wemple1977optical}. With PBE functional the value is 7.26, whereas with HSE06 the value is 5.11.}
\label{fig4}
\end{figure}
\textbf{Validation of functional}: The perpetual growth of modern computing architectures allows us to work on a method that goes beyond DFT for higher accuracy to understand the excited states. The method beyond DFT i.e., many body perturbation theory involves various steps in order to determine the optical properties~\cite{gali2012ab}. (i) The first step starts from basic DFT, that uses the energetics  and wavefunction obtained in DFT or hybrid exchange and correlation functional. (ii) Then, we perform the GW calculation on the top of DFT with appropriate functionals (viz. semi-local or hybrid). (iii) The final step is to include Bethe Salpeter Equation (BSE) on the top of GW. The presence of d electrons (transition metal atom Ti) in TiO$_2$ near fermi level are expected to be strongly correlated. This infers  the importance of electron and hole pair (excitons) in determining the optical properties. The inclusion of excitonic effect are well accomplished by using BSE on the top of GW. As mentioned in step (ii), that GW calculations are performed on the top of Kohn-Sham orbitals (i.e. DFT with either LDA, GGA (PBE) ~\cite{fuchs2005comparison,perdew1996generalized} or hybrid functional (HSE06)~\cite{heyd2003hybrid, krukau2008hybrid}). Therefore, it is crucial to choose a meaningful starting point for accurate estimation of the optical properties of (un)doped TiO$_2$. The paramount importance of exchange and co-relation functional arises for description of d-electron localization mainly in defective system. Further, the theoretical prediction of optical properties highly depends on the accuracy of real and imaginary part of dielectric function. We find that the calculation of real and imaginary part of a dielectric function from semi-local functional PBE produces a vast disagreement from experimental results. The results obtained from PBE functional strongly require the shifting of peaks manually in order to match the data from HSE06 as shown in Fig~\ref{fig4}. The static real part of the dielectric function from PBE is 7.26, whereas from HSE06 the value is 5.11 (see in Fig~\ref{fig4}b). The latter is more close to the experimental value of 5.62 ~\cite{wemple1977optical}. The reason  behind the discrepancy is incapability of PBE functional to determine the accurate defect levels and band gap. 
The HSE06 functional is expected to yield much improved properties such as correct symmetry of the ground state and a finite gap for semiconducting materials~\cite{deak2010accurate}. The important aspects of defect is certainty to locate the correct position of defect levels. This aspect can better be accomplished with the use of HSE06 functional. Therefore, we have proceeded with HSE06 functional as a starting point for our GW calculations. 
 Note that, despite GW calculations can reproduce accurate bandgap for many semiconductors and insulators, one term is still missing, which is the effect of ``hole" and electron-hole interaction.
Therefore, it is important to solve the four-point equation i.e., BSE that involves the propagation of two particles. 
 \subsection{Optical properties of pristine TiO$_2$}
 
\begin{figure}
 	\includegraphics[width=0.5\textwidth]{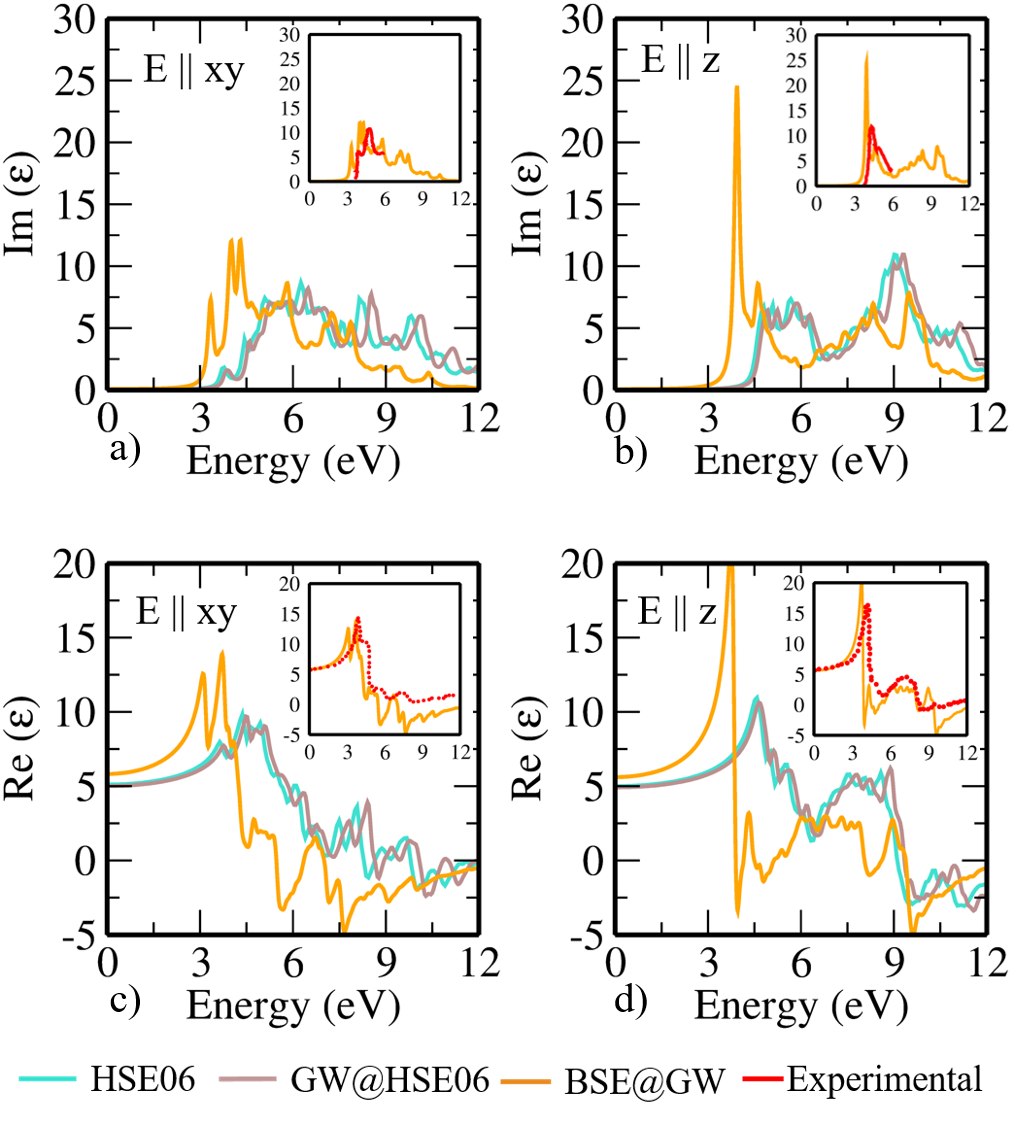} 
 	\caption{The imaginary part of dielectric function (a) along E$||$xy (b) along E$||$z and the real part of dielectric function (c) along  E$||$xy (d) along E$||$z of TiO$_2$ anatase. The experimental spectrum (red in color) is also shown for comparison purposes taken from Ref.~\cite{gonzalez1997infrared,wemple1977optical,hosaka1997optical}.}
 	\label{fig5}
 \end{figure}

  TiO$_2$ has tetragonal structure and it shows optical anisotropy. Therefore, it is important to understand the optical properties of TiO$_2$ in both the directions E$||$xy (i.e.; in-plane) and E$||$z (out-of-plane)~\cite{chiodo2010self,kang2010quasiparticle}.
  The imaginary and real part of the dielectric function of TiO$_2$ are calculated along E$||$xy and E$||$z direction as shown in Fig~\ref{fig5}. It is clearly observable from Fig~\ref{fig5}, that shape of spectra of GW@HSE06 and HSE06 are almost similar as both take into account the effect of screened coulomb interaction. However the spectra of BSE is quite different, which is assumed to be the reason for taking into consideration of electron-hole interaction. In the in-plane and the out-of-plane direction, we have found the initial peaks obtained from BSE@GW are very sharp, whereas in GW@HSE06 and HSE06 the initial peaks are not very sharp. This indicates us the excitonic contribution is non-negligible. This may be the reason that BSE@GW gives the correct oscillator strength, which indeed leads to a definite improvement of the result over GW@HSE06. 
The imaginary part of the dielectric function is defined as optical absorption. The onset of absorption starts at $\sim$ 3.0 eV from BSE calculations. The static value of real part of dielectric function (as shown in Fig~\ref{fig5}(c) and (d) along E$||$xy and E$||$z) are 5.80 and 5.61, which agrees well with the experimental data (of 5.82 and 5.62~\cite{gonzalez1997infrared,wemple1977optical} respectively). Our results show that along E$||$xy, the allowed optical transitions are lying at 3.45 eV, 4.04 eV and 4.28 eV corresponds to slight difference in optical absorption peaks from experimental value as shown in Fig~\ref{fig5}(a) and (b) insets~\cite{hosaka1997optical}. However, the nature of the absorption pattern is same. This discrepancy may be arisen, because in the experiments, higher order indirect process like scattering with a phonon are also involved. Now, along z direction (i.e.; E$||$z, out-of-plane) the optically allowed transitions are 4.01 eV and 4.7 eV. These type of spectra are important as they provide valuable information on optical transition probability corresponding to specific light wavelength. The other optical properties including refractive index ($\eta$), extinction coefficient ($\kappa$), reflectivity ($\textit{R}$), absorption coefficient ($\alpha$), optical conductivity ($\sigma$) and loss spectrum ($\textit{L}$) can be calculated once we know the real and imaginary part of the dielectric function. Following are the expressions used to determine the optical properties:
  \begin{align}
  	\eta = \frac {1}{\sqrt{2}} [{\sqrt{\textrm{Re}(\varepsilon)^2 + \textrm{Im}(\varepsilon)^2} + \textrm{Re}{(\varepsilon)}}]^\frac{1}{2}
  	\label{eqn2}
  \end{align}
  \begin{align}
  \kappa = \frac {1}{\sqrt{2}} [{\sqrt{\textrm{Re}(\varepsilon)^2 + \textrm{Im}(\varepsilon)^2} - \textrm{Re}{(\varepsilon)}}]^\frac{1}{2}
  \label{eqn3}
  \end{align}
  \begin{align}
  R = \dfrac{[\eta - 1]^2 + {\kappa}^2}{[\eta + 1]^2 + {\kappa}^2}
  \label{eqn4}
  \end{align}
  \begin{align}
  	\alpha = \dfrac{2 \omega \kappa} {c}
  	\label{eqn5}
  \end{align}
   \begin{align}
   \sigma = \dfrac {\alpha \eta c} {4\pi}
   \label{eqn6}
   \end{align}
   \begin{align}
   L = \dfrac{\textrm{Im}{(\varepsilon)}}{\textrm{Re}(\varepsilon)^2 + \textrm{Im}(\varepsilon)^2}
   \label{eqn7}
   \end{align}
   
   \begin{figure}
   	\includegraphics[width=0.5\textwidth]{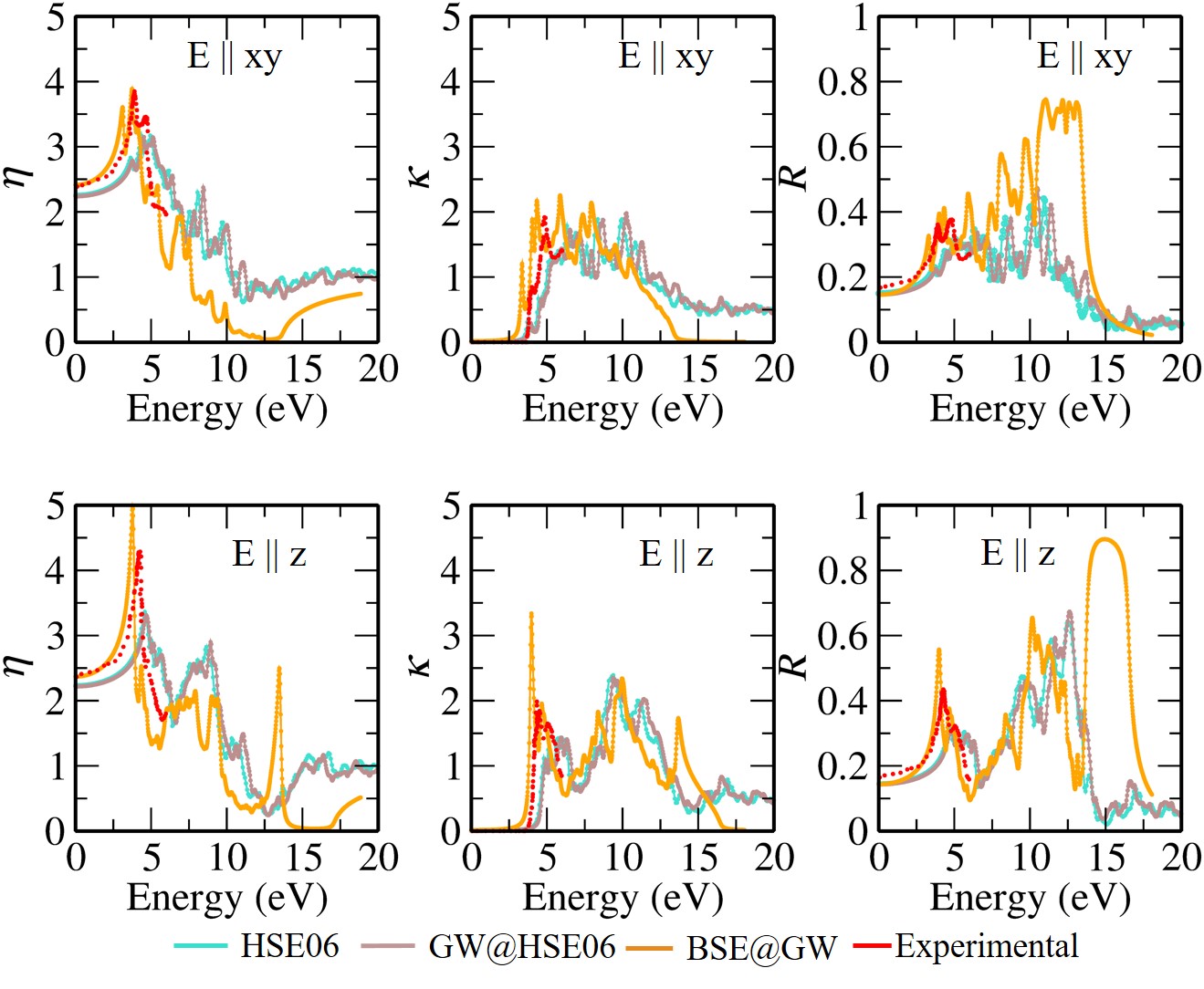} 
   	\caption{Optical properties of TiO$_2$ anatase: (a) refractive index ($\eta$), (b) extinction coefficient ($\kappa$), (c) reflectivity ($\textit{R}$) along E$||$xy direction and (d) refractive index ($\eta$), (e) extinction coefficient ($\kappa$), (f) reflectivity ($\textit{R}$) along E$||$z direction.}
   	\label{fig6}
   \end{figure} 
   \begin{figure}
   	\includegraphics[width=0.5\textwidth]{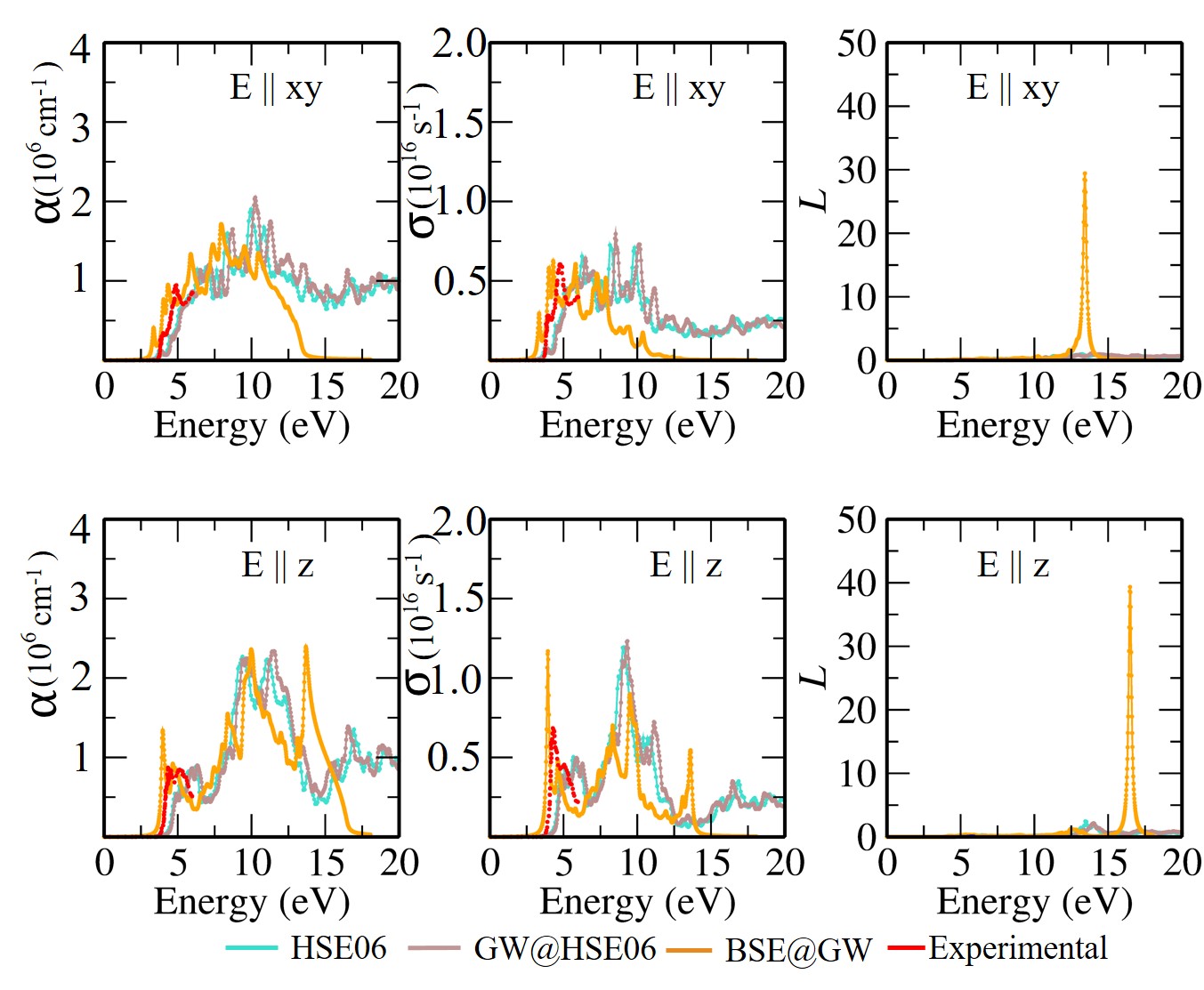} 
   	\caption{(a) absorption coefficient ($\alpha$), (b) optical conductivity ($\sigma$), (c) loss spectrum ($\textit{L}$) for electric field polarized along xy direction and (d) absorption coefficient ($\alpha$), (e) optical conductivity ($\sigma$), (f) loss spectrum ($\textit{L}$) along z direction.}
   	\label{fig7}
   \end{figure}
The optical properties of TiO$_2$ anatase in both the direction E$||$xy and E$||$z are given in Fig~\ref{fig6} and Fig~\ref{fig7}. The optical properties are shown in energy range from 0 to 20 eV. The static refractive index calculated for E$||$xy is 2.41 and E$||$z is 2.36. These values are close in agreement with the experimental results~\cite{gonzalez1997infrared}. The experimental data is shown for 0-6 eV range. It's therefore validated that BSE calculations are good enough to reproduce the undoped TiO$_2$ case. We therefore have used this method to study the optical properties of doped TiO$_2$ system. 

 \subsection{Optical properties of doped TiO$_2$}
 The importance of many body perturbation theory to study n-type dopants in TiO$_2$ rutile is recently reported~\cite{atambo2019electronic}. It unveils that the presence of defects can significantly affect the optical properties of TiO$_2$ rutile. Here, we have doped different non-metals X = (S, Se, N and C) in TiO$_2$ anatase, where X substituted at oxygen site X$_\textrm{O}$ to enhance their optical properties. We have already shown~\cite{PB} that these configurations are most stable at various environmental conditions (viz. O-rich, O-poor, O-intermediate) because of their lower formation energy.
 \begin{figure}
 	\includegraphics[width=0.5\textwidth]{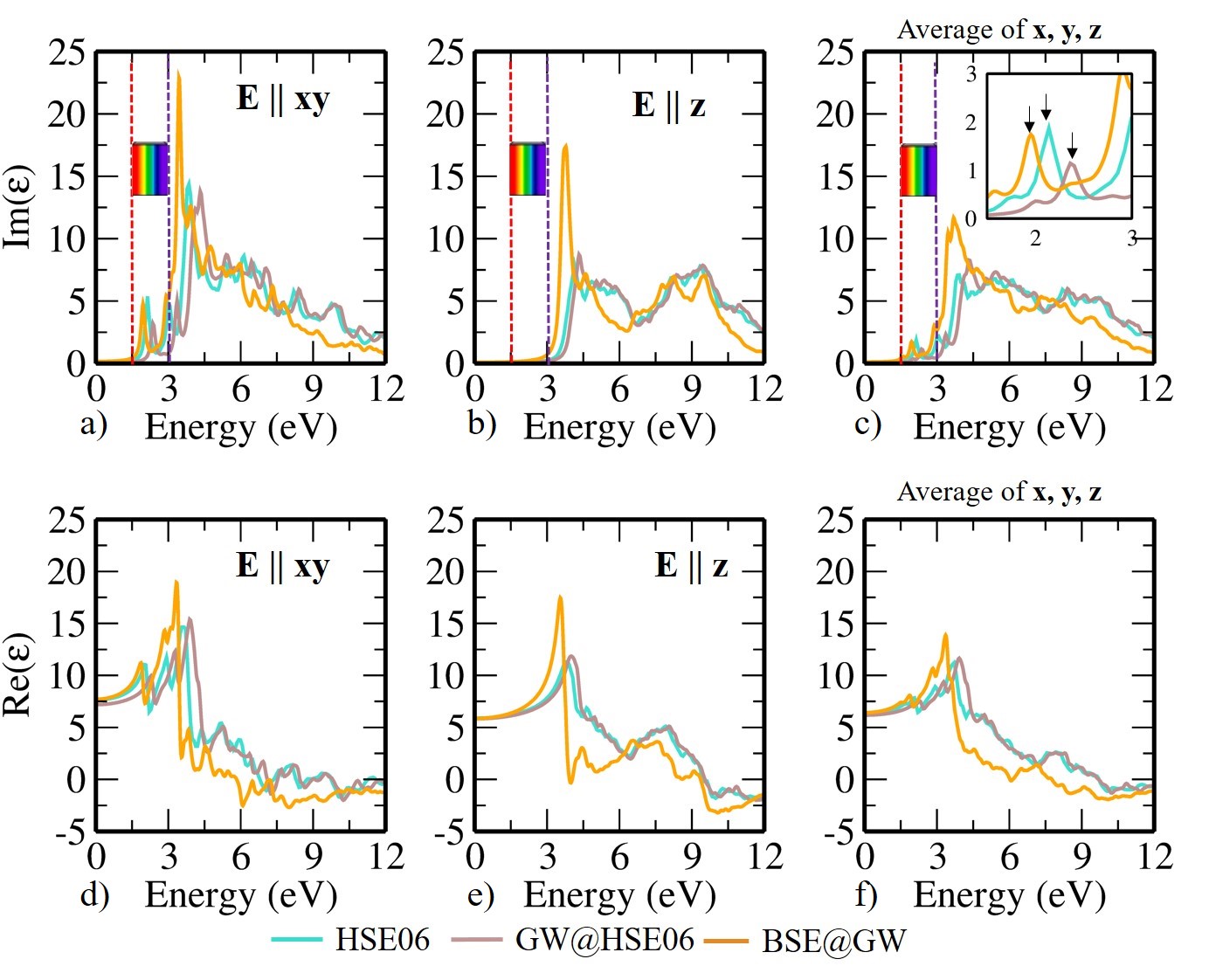} 
 	\caption{ Imaginary part of the dielectric function (a) along E$||$xy (b) along E$||$z (c) average along x, y and z directions and the real part of the dielectric function  (d) along E$||$xy (e) along E$||$z and (f) average along x, y and z directions for S doped TiO$_2$ anatase.}
 	\label{fig8}
 \end{figure}
 
\textbf{S doped TiO$_2$:} From Fig~\ref{fig5}(a) and (b), we have observed that, there is no absorption peak below 3 eV for pristine TiO$_2$. However, for S doped TiO$_2$ (S$_\textrm{O}$), the absorption peaks are lying at lower energy below 3 eV (see Fig~\ref{fig8}(a)). Pristine TiO$_2$ anatase exhibits a significant optical anisotropy, associated with its tetragonal symmetry. Hence, the presence of anisotropy is also obvious for the doped systems, and it is more pronounced in the low energy part of the spectra. It is interesting to note that for S$_\textrm{O}$, the E$||$xy is optically active for light polarization, whereas E$||$z has no contribution in energy range below 3 eV. The average of all the polarization direction x, y and z is shown in Fig~\ref{fig8}(c). The inset shows the absorption spectra in the visible range (1.5 eV to 3.0 eV). The black arrows in the inset indicate the position of first absorption peak in the visible region. The HSE06 functional gives the absorption peak at 2.2 eV. The GW on the top of HSE06 functional gives the peak at 2.4 eV. Finally, BSE gives the peak value at 1.98 eV. The calculated first absorption peaks from HSE06, GW@HSE06 and BSE@GW are lying in the visible region. Here, we also note that the shift in BSE spectra with respect to GW is less as compared to shift observed in pristine anatase TiO$_2$.  The reason may be increased screening due to the excess electrons of the dopants. Therefore, Sulphur doped TiO$_2$ is useful for the applications in photo-active devices. Since, we know the imaginary  and real part of the dielectric function from Fig~\ref{fig8}, the other optical properties can be determined using Eq.~\ref{eqn2}-7.\\
\begin{figure}
	\includegraphics[width=0.5\textwidth]{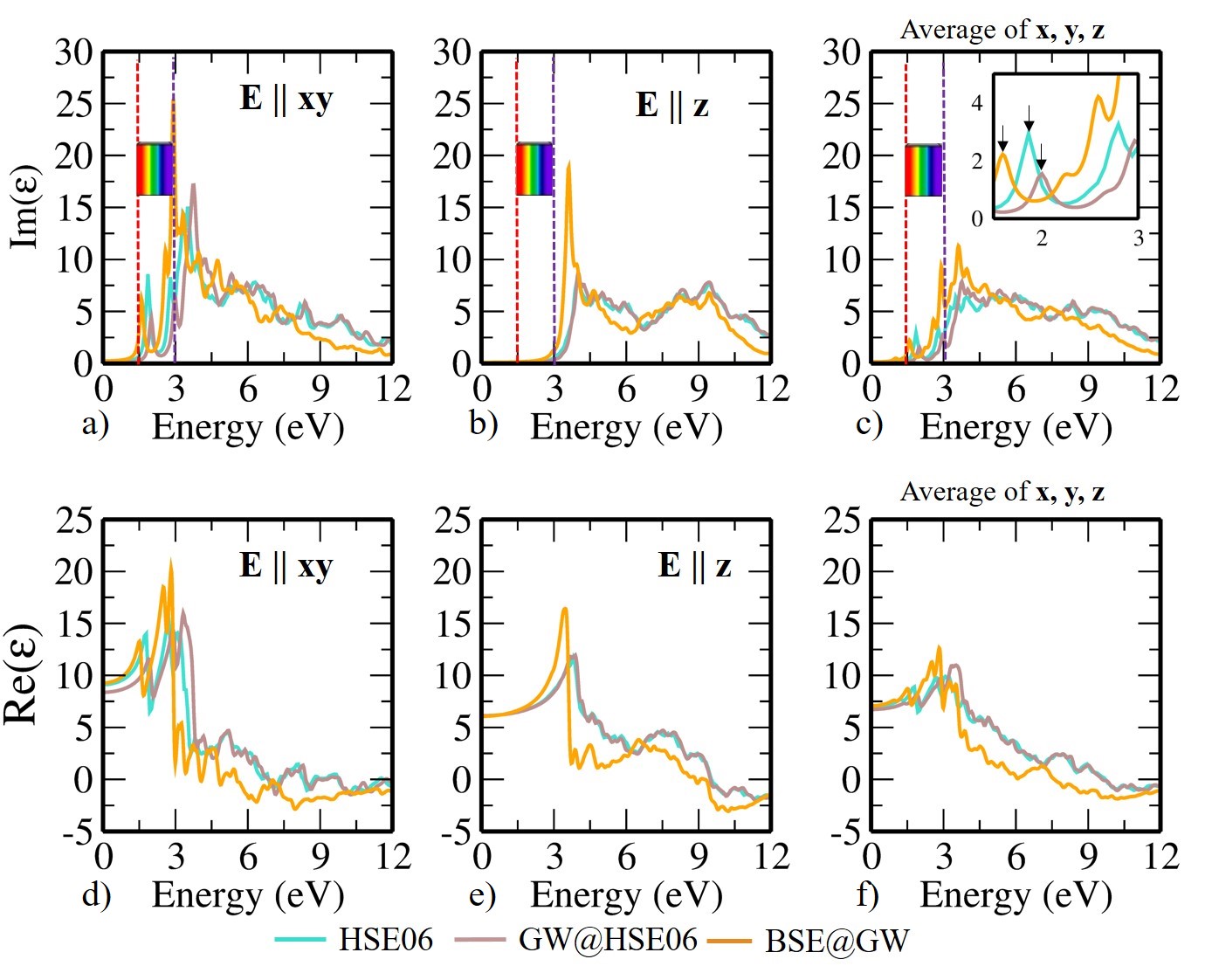} 
	\caption{ For Se doped TiO$_2$ anatase: Imaginary part of the dielectric function (a) along E$||$xy (b) along E$||$z (c) average along x, y and z directions and the real part of the dielectric function  (d) along E$||$xy (e) along E$||$z and (f) average along x, y and z directions.}
	\label{fig9}
\end{figure}
\textbf{Se doped TiO$_2$:}
Similarly, for Se doped system (Se$_\textrm{O}$), the absorption peak below 3 eV is obtained only along E$||$xy direction shown in Fig~\ref{fig9}(a). There is no significant contribution at low energy region for the case of electric field polarized along z direction (see Fig~\ref{fig9}(b)). The average of x, y and z polarisation are given in Fig~\ref{fig9}(c). The inset shows the absorption peaks in the visible region range. The peak positions in inset are marked with black arrows. The HSE06 and GW@HSE06 follows the same peak pattern. The absorption peak value for HSE06, GW@HSE06 and BSE@HSE06 are 1.91 eV, 2.02eV and 1.63 eV. Again the values lying in the visible region range. Hence, Se doped TiO$_2$ is useful for applications in opto-electronic and photo-active devices. 
\\
 \textbf{N doped TiO$_2$:}
 \begin{figure}
		\includegraphics[width=0.5\textwidth]{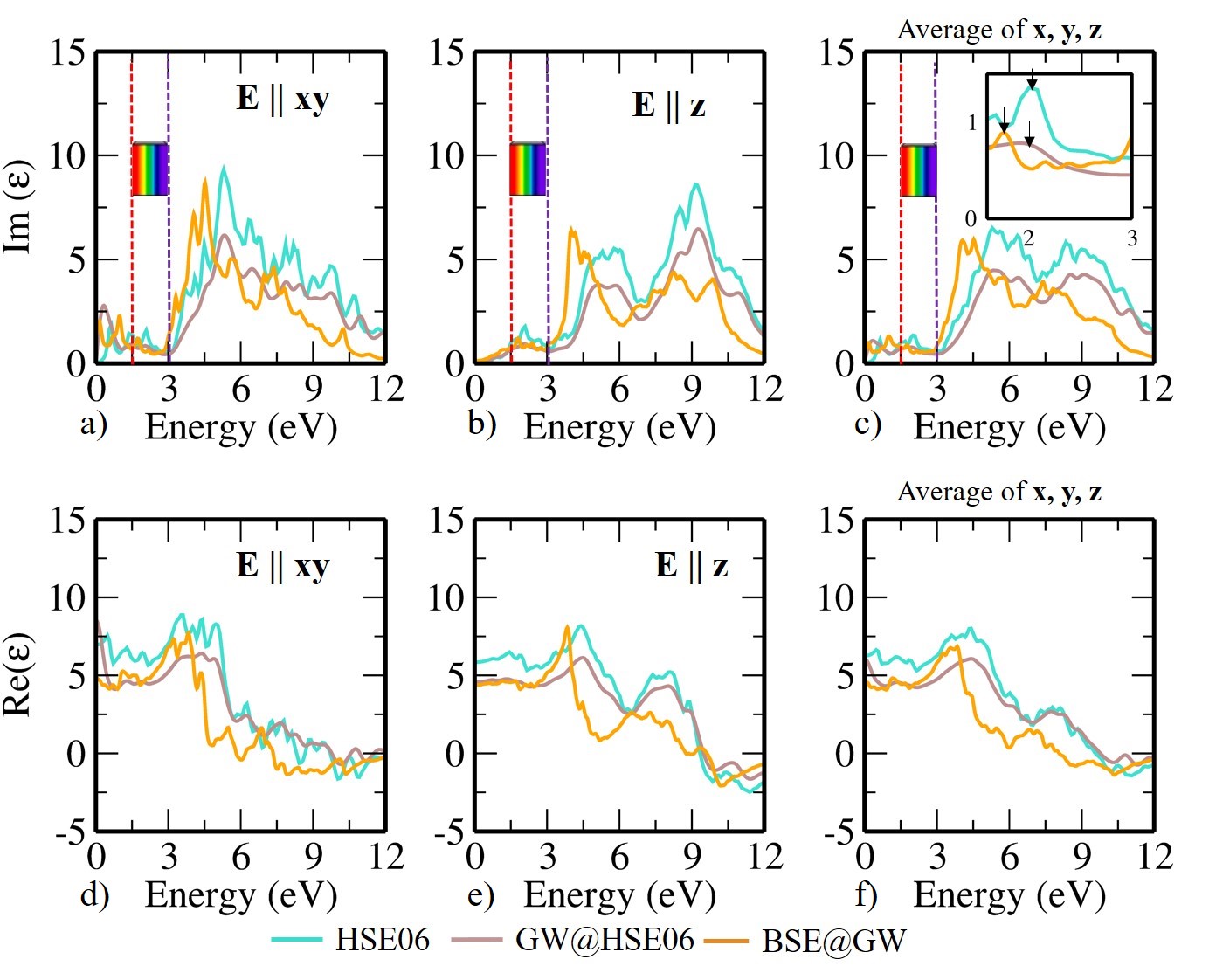} 
	\caption{For N doped TiO$_2$ anatase: (a) along E$||$xy (b) along E$||$z (c) average along x, y and z directions show the absorption spectra i.e., imaginary part of the dielectric function and (d) along E$||$xy (e) along E$||$z and (f) average along x, y and z directions are real part of the dielectric function.}
	\label{fig10}
\end{figure}
The absorption peaks at low energy below 3eV are observed for E$||$xy as well as for E$||$z direction as shown in Fig~\ref{fig10}. This means, unlike the case of S and Se doped TiO$_2$, both the directions are optically active for polarization of light. Here, we have found the slight shift in spectra shape of HSE06 with respect to GW@HSE06. This shift may be orginated because of the smaller size of N atom as compared to S and Se. The shielding effect tends to increase as the atomic radius increases, because a new shell of electrons is added up. The screening effect in the case of N doped system is less than S and Se case. The inset  shows the visible region range, where black arrows represent those peaks that are in visible range. The absorption peak values for HSE06, GW@HSE06 and BSE@HSE06 are respectively 2.04 eV, 2.02eV and 1.80 eV.\\ 
\textbf{C doped TiO$_2$:}
Like N, the absorption peaks at low energy below 3eV are observed for E$||$xy as well as for E$||$z direction as shown in Fig~\ref{fig11}. Hence, they are optically active for both the direction. From Fig~\ref{fig11}(a), we see that peak near-infrared region at low energy range are more intense and sharper as compared to peak at higher energy range. On the contrary, along z direction (see Fig~\ref{fig11}(b)) the peaks are not very intense. Therefore, the  major optical contribution in C doped TiO$_2$ is along E$||$xy direction. The average of all the directions is shown in Fig~\ref{fig11}(c). The black arrows in the inset indicate only those absorption peaks that are lying in the visible region. The absorption peak values for HSE06, GW@HSE06 and BSE@HSE06 are respectively 1.7 eV, 1.6 eV and 1.5 eV. The absorption peaks are red shifted in the visible region. Note that, in the case of N and C doped TiO$_2$ high energy region is significantly affected on comparing it with pristine TiO$_2$, whereas such behavior is not observed for S and Se dopants.
\begin{figure}
	\includegraphics[width=0.5\textwidth]{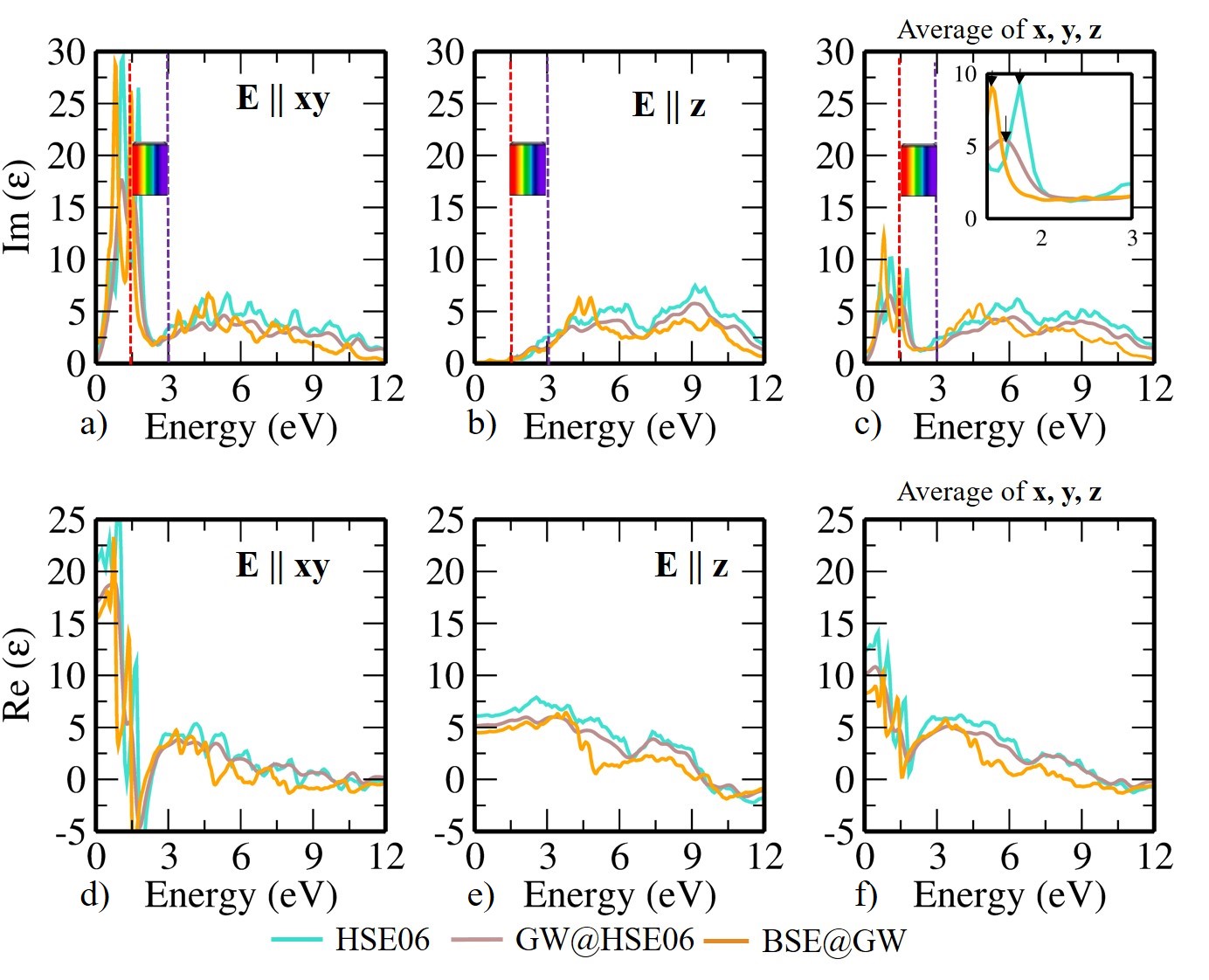} 
	\caption{Imaginary and real part of the dielectric function for C doped TiO$_2$ anatase	for light polarized along E $||$ xy, E $||$ z and average of x, y and z directions.}
	\label{fig11}
\end{figure}\\
\begin{figure}
	\includegraphics[width=0.5\textwidth]{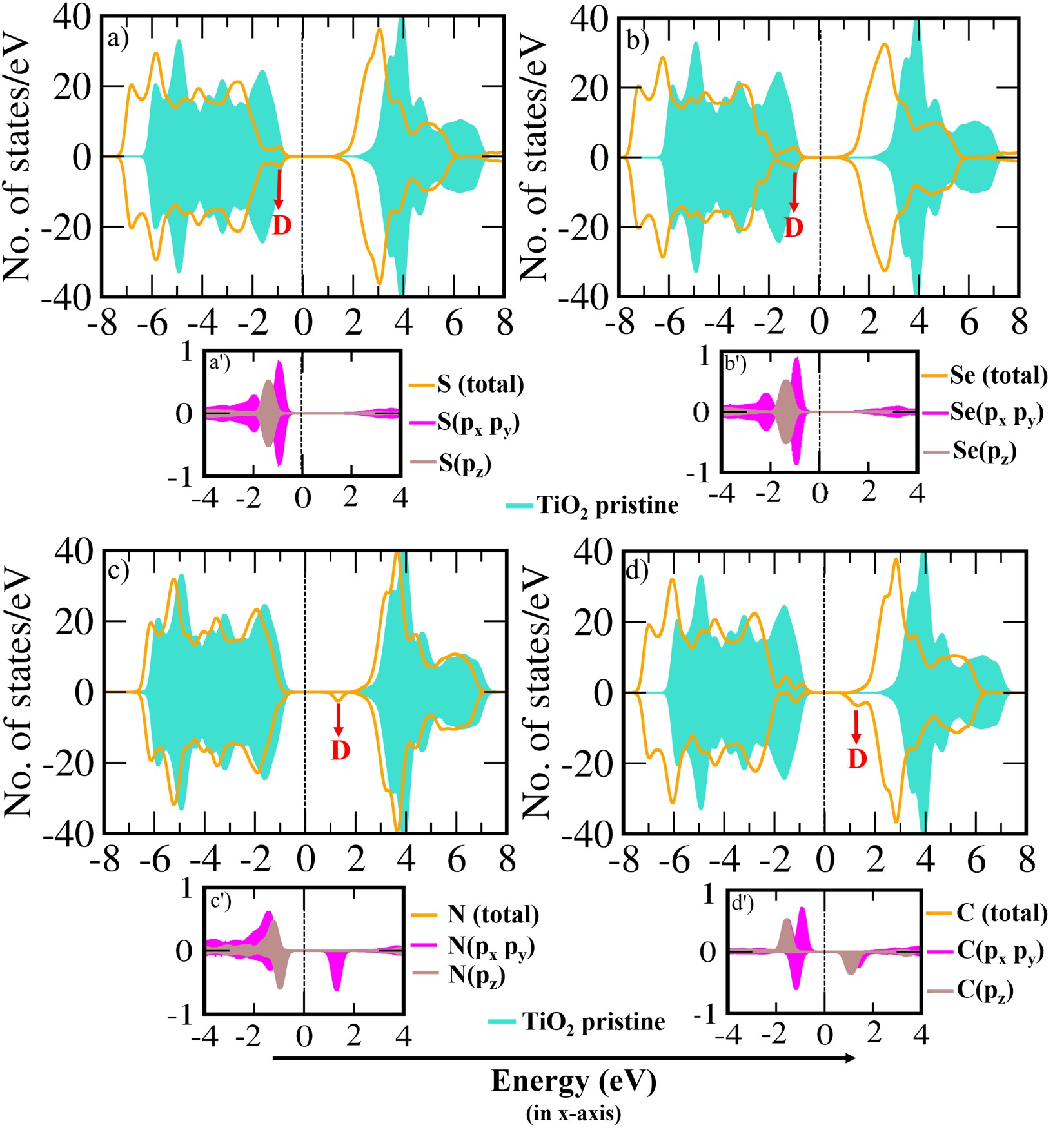} 
	\caption{ a) DOS and a$'$) PDOS of S doped TiO{$_2$} (contribution of p orbitals  of S along x, y and z direction. b) DOS and b$'$) PDOS of Se doped TiO{$_2$}. c) DOS and c$'$) PDOS of N doped TiO{$_2$}. d)DOS and d$'$) PDOS of C doped TiO{$_2$}. The red arrow represent the defect peaks. Note that, we have plotted the DOS for both the systems (pristine and doped TiO$_2$) by shifting the Fermi level  at 0 as our reference point.}
	\label{fig12}
\end{figure}
\begin{figure}
	\includegraphics[width=0.5\textwidth]{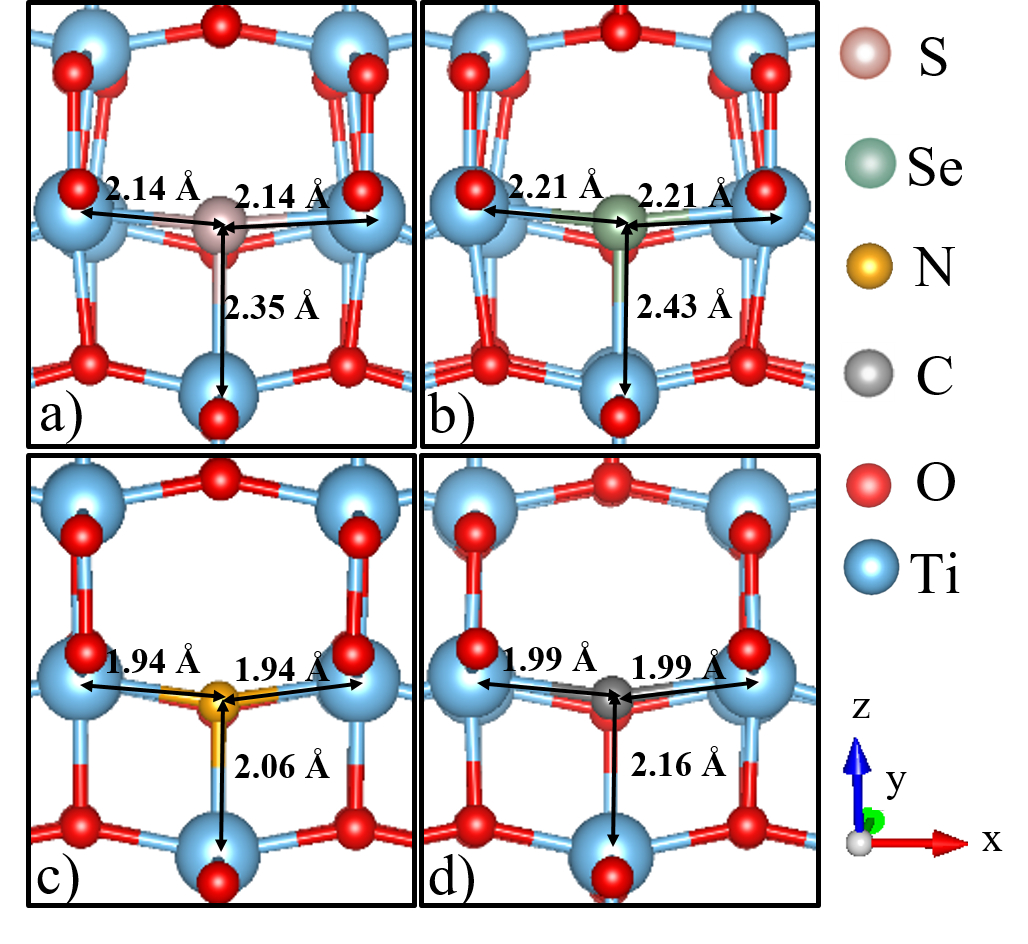} 
	\caption{Optimised geometry of a) S doped TiO$_2$. b) Se doped TiO$_2$. c) N doped TiO$_2$. d) C doped TiO$_2$. Bond lengths (in \AA) for doped anatase TiO$_2$ in xy and z directions are shown in black arrows.}
	\label{fig13}
\end{figure}
From the above discussion we have concluded some important point for X$_\textrm{O}$ in TiO$_2$ anatase: (i) In low energy range (0-3 eV), S and Se dopants both are optically active only in xy direction. In order to know the underlying reason for this optically active (inactive) behavior in low energy region, we have explicitly plotted the density of states (DOS) and partial density of states (PDOS) as shown in Fig~\ref{fig12}. Note that, DOS of doped TiO$_2$ is shifted from pristine TiO$_2$ (see Fig~\ref{fig12}), in order to align the Fermi level of doped system at 0 as our common reference point. One can also visualize the same by having two Fermi-levels of the pristine and doped system. In that case, the DOS of two systems will be completely overlapped (excluding the extra defect level as in the defected system). From DOS, we can say that S and Se dopants in TiO$_2$ are p-type in nature, as defect peaks are lying near valence band maxima (VBM) as represented by D (see Fig~\ref{fig12}(a) and (b)). The inset is showing the contribution of p$_x$p$_y$ orbitals (of S/Se) that are being more dominated towards the Fermi level (see Fig~\ref{fig12}(a$'$)and (b$'$)). Hence, initial optical transitions will be preferable by p$_x$p$_y$ orbitals in low energy region. This also explains the reason that the z-direction is optically inactive in low energy region. Afterwards, in high energy region, both p$_x$p$_y$ and p$_z$ orbital can contribute. In short, p-type dopants are optically active for light polarized along xy direction in TiO$_2$ anatase for low energy region.

(ii) N and C dopants are optically active along both xy and z direction in low energy regime. From DOS, the defect peaks are observed near VBM as well as near conduction band minimum (CBm). We have already shown in our previous findings that  N and C dopants have a tendency to accept the electron (prefer negative charges)~\cite{PB}. This is possible only when they have unoccupied states near CBm. Thus, we can say that states near CBm play a significant contribution rather than states near VBM. Therefore, defect peak represented by D is shown near CBm. Hence, N and C are n-type dopants as shown in Fig~\ref{fig12}(c) and (d). The inset is showing the individual contribution of N/C p$_x$p$_y$ and p$_z$ orbitals (see Fig~\ref{fig12}(c$'$)and (d$'$)). This shows p$_x$p$_y$ and p$_z$ orbitals are actively participated in initial optical transitions in low energy region. Therefore, n-type dopants are optically active for light polarized along xy and z direction in low energy region.

Further, we have measured the bond length to show the distortion in structure caused by doping. For S and Se dopants, the stretching along xy and z direction is highly significant with respect to TiO$_2$ pristine (see Table~\ref{Table1}). The distortion is more along z direction. For the case of N and C dopants, only the slight distortion is observed as shown in Fig~\ref{fig13}. This shows the presence of anisotropy in doped systems. In addition, this distortion may also play a role to make the system optically active (inactive) in low energy region.

\begin{table}[htbp]
	\caption{Bond length of various dopants along xy and z direction} 
	\begin{center}
		\begin{tabular}[c]{|c|r|c|r|} \hline
			Bond length & along xy (\AA) & along z (\AA)\\ \hline
			O-Ti   & 1.92  & 1.99 \\ \hline
			S-Ti   & 2.14  & 2.35  \\ \hline
			Se-Ti  & 2.21  & 2.43  \\ \hline
			N-Ti   & 1.94  & 2.06  \\ \hline
			C-Ti   & 1.99  & 2.16    
			\\ \hline 
		\end{tabular}
		\label{Table1}
	\end{center}
\end{table}

 \section{Conclusions}
 In summary, we have presented an exhaustive study to understand the optical properties of pristine and non-metal doped TiO$_2$ anatase using CC diagram and many body perturbation theory approaches. As a first step, we have validated the number of occupied (NO) and unoccupied bands (NV) requisite for BSE convergence. Further, we have validated the initial starting point for the single shot GW calculations. The good choice of initial starting point (exchange and co-relation functional) is important for description of d-electron localization especially in defective systems. The CC diagram is only limited to obtain the optical absorption and emission peaks (not applicable for pristine TiO$_2$). The dopants (X) significantly perturb the system of TiO$_2$ anatase and therefore, results in complicated spectrum of quasi-local modes. The broadening of the spectra has not been considered in the CC diagram. Therefore, it is not able to capture the fine spectra, which is observed in experiments. Hence, we have proceeded towards the MBPT approach that includes the excitonic effects (BSE) and capable to transpire the fine optical spectra close to experiments. The anisotropy in non-metal doped TiO$_2$ anatase plays an important role in the optical spectra.  Optical excitations are found to be strongly anisotropic. We have obtained that in low energy region, n-type doped TiO$_2$ anatase is optically active in both x, z direction, whereas the p-type doped TiO$_2$ anatase is optically inactive z-direction.  We have found that in all the doped systems optically allowed transitions are introduced well below 3 eV, thus improving its opto-electronic and solar absorption properties.

\section{Acknowledgement}
PB acknowledges UGC, India, for the senior research fellowship [grant no. 20/12/2015 (ii) EUV]. SS acknowledges CSIR, India, for the senior research fellowship [grant no. 09/086 (1231) 2015-EMR-I]. SB acknowledge the financial support from YSS-SERB research grant, SERB India (grant no. YSS/2015/001209). We acknowledge the High Performance Computing (HPC) facility at IIT Delhi for computational resources.

     

\begin{thebibliography}{64}
\expandafter\ifx\csname natexlab\endcsname\relax\def\natexlab#1{#1}\fi
\expandafter\ifx\csname bibnamefont\endcsname\relax
  \def\bibnamefont#1{#1}\fi
\expandafter\ifx\csname bibfnamefont\endcsname\relax
  \def\bibfnamefont#1{#1}\fi
\expandafter\ifx\csname citenamefont\endcsname\relax
  \def\citenamefont#1{#1}\fi
\expandafter\ifx\csname url\endcsname\relax
  \def\url#1{\texttt{#1}}\fi
\expandafter\ifx\csname urlprefix\endcsname\relax\def\urlprefix{URL }\fi
\providecommand{\bibinfo}[2]{#2}
\providecommand{\eprint}[2][]{\url{#2}}

\bibitem[{\citenamefont{Chen et~al.}(2014)\citenamefont{Chen, Yu, Han, Yan,
  Liu, Zhang, Zhang, Yu, and Long}}]{chen2014electrospun}
\bibinfo{author}{\bibfnamefont{S.}~\bibnamefont{Chen}},
  \bibinfo{author}{\bibfnamefont{M.}~\bibnamefont{Yu}},
  \bibinfo{author}{\bibfnamefont{W.-P.} \bibnamefont{Han}},
  \bibinfo{author}{\bibfnamefont{X.}~\bibnamefont{Yan}},
  \bibinfo{author}{\bibfnamefont{Y.-C.} \bibnamefont{Liu}},
  \bibinfo{author}{\bibfnamefont{J.-C.} \bibnamefont{Zhang}},
  \bibinfo{author}{\bibfnamefont{H.-D.} \bibnamefont{Zhang}},
  \bibinfo{author}{\bibfnamefont{G.-F.} \bibnamefont{Yu}}, \bibnamefont{and}
  \bibinfo{author}{\bibfnamefont{Y.-Z.} \bibnamefont{Long}},
  \bibinfo{journal}{RSC Advances} \textbf{\bibinfo{volume}{4}},
  \bibinfo{pages}{46152} (\bibinfo{year}{2014}).

\bibitem[{\citenamefont{Bai et~al.}(2014)\citenamefont{Bai, Mora-Sero,
  De~Angelis, Bisquert, and Wang}}]{bai2014titanium}
\bibinfo{author}{\bibfnamefont{Y.}~\bibnamefont{Bai}},
  \bibinfo{author}{\bibfnamefont{I.}~\bibnamefont{Mora-Sero}},
  \bibinfo{author}{\bibfnamefont{F.}~\bibnamefont{De~Angelis}},
  \bibinfo{author}{\bibfnamefont{J.}~\bibnamefont{Bisquert}}, \bibnamefont{and}
  \bibinfo{author}{\bibfnamefont{P.}~\bibnamefont{Wang}},
  \bibinfo{journal}{Chemical reviews} \textbf{\bibinfo{volume}{114}},
  \bibinfo{pages}{10095} (\bibinfo{year}{2014}).

\bibitem[{\citenamefont{Gr{\"a}tzel}(2005)}]{gratzel2005solar}
\bibinfo{author}{\bibfnamefont{M.}~\bibnamefont{Gr{\"a}tzel}},
  \bibinfo{journal}{Inorganic chemistry} \textbf{\bibinfo{volume}{44}},
  \bibinfo{pages}{6841} (\bibinfo{year}{2005}).

\bibitem[{\citenamefont{O'regan and Gr{\"a}tzel}(1991)}]{o1991low}
\bibinfo{author}{\bibfnamefont{B.}~\bibnamefont{O'regan}} \bibnamefont{and}
  \bibinfo{author}{\bibfnamefont{M.}~\bibnamefont{Gr{\"a}tzel}},
  \bibinfo{journal}{nature} \textbf{\bibinfo{volume}{353}},
  \bibinfo{pages}{737} (\bibinfo{year}{1991}).

\bibitem[{\citenamefont{Ghosh and English}(2012)}]{ghosh2012ab}
\bibinfo{author}{\bibfnamefont{S.}~\bibnamefont{Ghosh}} \bibnamefont{and}
  \bibinfo{author}{\bibfnamefont{N.~J.} \bibnamefont{English}},
  \bibinfo{journal}{Physical Review B} \textbf{\bibinfo{volume}{86}},
  \bibinfo{pages}{235203} (\bibinfo{year}{2012}).

\bibitem[{\citenamefont{Kim et~al.}(2004)\citenamefont{Kim, Kim, Kim, Hwang,
  and Jeong}}]{kim2004high}
\bibinfo{author}{\bibfnamefont{S.~K.} \bibnamefont{Kim}},
  \bibinfo{author}{\bibfnamefont{W.-D.} \bibnamefont{Kim}},
  \bibinfo{author}{\bibfnamefont{K.-M.} \bibnamefont{Kim}},
  \bibinfo{author}{\bibfnamefont{C.~S.} \bibnamefont{Hwang}}, \bibnamefont{and}
  \bibinfo{author}{\bibfnamefont{J.}~\bibnamefont{Jeong}},
  \bibinfo{journal}{Applied Physics Letters} \textbf{\bibinfo{volume}{85}},
  \bibinfo{pages}{4112} (\bibinfo{year}{2004}).

\bibitem[{\citenamefont{Liu et~al.}(2013)\citenamefont{Liu, Chen, Liu, Andrews,
  Hahn, and Yang}}]{liu2013large}
\bibinfo{author}{\bibfnamefont{B.}~\bibnamefont{Liu}},
  \bibinfo{author}{\bibfnamefont{H.~M.} \bibnamefont{Chen}},
  \bibinfo{author}{\bibfnamefont{C.}~\bibnamefont{Liu}},
  \bibinfo{author}{\bibfnamefont{S.~C.} \bibnamefont{Andrews}},
  \bibinfo{author}{\bibfnamefont{C.}~\bibnamefont{Hahn}}, \bibnamefont{and}
  \bibinfo{author}{\bibfnamefont{P.}~\bibnamefont{Yang}},
  \bibinfo{journal}{Journal of the American Chemical Society}
  \textbf{\bibinfo{volume}{135}}, \bibinfo{pages}{9995} (\bibinfo{year}{2013}).

\bibitem[{\citenamefont{Wang et~al.}(2014{\natexlab{a}})\citenamefont{Wang,
  Zhang, Li, Li, and Lin}}]{wang2014first}
\bibinfo{author}{\bibfnamefont{Y.}~\bibnamefont{Wang}},
  \bibinfo{author}{\bibfnamefont{R.}~\bibnamefont{Zhang}},
  \bibinfo{author}{\bibfnamefont{J.}~\bibnamefont{Li}},
  \bibinfo{author}{\bibfnamefont{L.}~\bibnamefont{Li}}, \bibnamefont{and}
  \bibinfo{author}{\bibfnamefont{S.}~\bibnamefont{Lin}},
  \bibinfo{journal}{Nanoscale research letters} \textbf{\bibinfo{volume}{9}},
  \bibinfo{pages}{46} (\bibinfo{year}{2014}{\natexlab{a}}).

\bibitem[{\citenamefont{Janisch et~al.}(2005)\citenamefont{Janisch, Gopal, and
  Spaldin}}]{janisch2005transition}
\bibinfo{author}{\bibfnamefont{R.}~\bibnamefont{Janisch}},
  \bibinfo{author}{\bibfnamefont{P.}~\bibnamefont{Gopal}}, \bibnamefont{and}
  \bibinfo{author}{\bibfnamefont{N.~A.} \bibnamefont{Spaldin}},
  \bibinfo{journal}{Journal of Physics: Condensed Matter}
  \textbf{\bibinfo{volume}{17}}, \bibinfo{pages}{R657} (\bibinfo{year}{2005}).

\bibitem[{\citenamefont{Ribao et~al.}(2017)\citenamefont{Ribao, Rivero, and
  Ortiz}}]{ribao2017tio2}
\bibinfo{author}{\bibfnamefont{P.}~\bibnamefont{Ribao}},
  \bibinfo{author}{\bibfnamefont{M.~J.} \bibnamefont{Rivero}},
  \bibnamefont{and} \bibinfo{author}{\bibfnamefont{I.}~\bibnamefont{Ortiz}},
  \bibinfo{journal}{Environmental Science and Pollution Research}
  \textbf{\bibinfo{volume}{24}}, \bibinfo{pages}{12628} (\bibinfo{year}{2017}).

\bibitem[{\citenamefont{Ma et~al.}(2014)\citenamefont{Ma, Dai, Yu, and
  Huang}}]{ma2014noble}
\bibinfo{author}{\bibfnamefont{X.}~\bibnamefont{Ma}},
  \bibinfo{author}{\bibfnamefont{Y.}~\bibnamefont{Dai}},
  \bibinfo{author}{\bibfnamefont{L.}~\bibnamefont{Yu}}, \bibnamefont{and}
  \bibinfo{author}{\bibfnamefont{B.}~\bibnamefont{Huang}},
  \bibinfo{journal}{Scientific reports} \textbf{\bibinfo{volume}{4}}
  (\bibinfo{year}{2014}).

\bibitem[{\citenamefont{Zhu et~al.}(2009)\citenamefont{Zhu, Qiu, Iancu, Chen,
  Pan, Wang, Dimitrijevic, Rajh, Meyer~III, Paranthaman et~al.}}]{zhu2009band}
\bibinfo{author}{\bibfnamefont{W.}~\bibnamefont{Zhu}},
  \bibinfo{author}{\bibfnamefont{X.}~\bibnamefont{Qiu}},
  \bibinfo{author}{\bibfnamefont{V.}~\bibnamefont{Iancu}},
  \bibinfo{author}{\bibfnamefont{X.-Q.} \bibnamefont{Chen}},
  \bibinfo{author}{\bibfnamefont{H.}~\bibnamefont{Pan}},
  \bibinfo{author}{\bibfnamefont{W.}~\bibnamefont{Wang}},
  \bibinfo{author}{\bibfnamefont{N.~M.} \bibnamefont{Dimitrijevic}},
  \bibinfo{author}{\bibfnamefont{T.}~\bibnamefont{Rajh}},
  \bibinfo{author}{\bibfnamefont{H.~M.} \bibnamefont{Meyer~III}},
  \bibinfo{author}{\bibfnamefont{M.~P.} \bibnamefont{Paranthaman}},
  \bibnamefont{et~al.}, \bibinfo{journal}{Physical review letters}
  \textbf{\bibinfo{volume}{103}}, \bibinfo{pages}{226401}
  (\bibinfo{year}{2009}).

\bibitem[{\citenamefont{Wang et~al.}(2014{\natexlab{b}})\citenamefont{Wang,
  Sun, Huang, Li, and Yang}}]{wang2014band}
\bibinfo{author}{\bibfnamefont{J.}~\bibnamefont{Wang}},
  \bibinfo{author}{\bibfnamefont{H.}~\bibnamefont{Sun}},
  \bibinfo{author}{\bibfnamefont{J.}~\bibnamefont{Huang}},
  \bibinfo{author}{\bibfnamefont{Q.}~\bibnamefont{Li}}, \bibnamefont{and}
  \bibinfo{author}{\bibfnamefont{J.}~\bibnamefont{Yang}}, \bibinfo{journal}{The
  Journal of Physical Chemistry C} \textbf{\bibinfo{volume}{118}},
  \bibinfo{pages}{7451} (\bibinfo{year}{2014}{\natexlab{b}}).

\bibitem[{\citenamefont{Pelaez et~al.}(2012)\citenamefont{Pelaez, Nolan,
  Pillai, Seery, Falaras, Kontos, Dunlop, Hamilton, Byrne, O'shea
  et~al.}}]{pelaez2012review}
\bibinfo{author}{\bibfnamefont{M.}~\bibnamefont{Pelaez}},
  \bibinfo{author}{\bibfnamefont{N.~T.} \bibnamefont{Nolan}},
  \bibinfo{author}{\bibfnamefont{S.~C.} \bibnamefont{Pillai}},
  \bibinfo{author}{\bibfnamefont{M.~K.} \bibnamefont{Seery}},
  \bibinfo{author}{\bibfnamefont{P.}~\bibnamefont{Falaras}},
  \bibinfo{author}{\bibfnamefont{A.~G.} \bibnamefont{Kontos}},
  \bibinfo{author}{\bibfnamefont{P.~S.} \bibnamefont{Dunlop}},
  \bibinfo{author}{\bibfnamefont{J.~W.} \bibnamefont{Hamilton}},
  \bibinfo{author}{\bibfnamefont{J.~A.} \bibnamefont{Byrne}},
  \bibinfo{author}{\bibfnamefont{K.}~\bibnamefont{O'shea}},
  \bibnamefont{et~al.}, \bibinfo{journal}{Applied Catalysis B: Environmental}
  \textbf{\bibinfo{volume}{125}}, \bibinfo{pages}{331} (\bibinfo{year}{2012}).

\bibitem[{\citenamefont{Irie et~al.}(2003{\natexlab{a}})\citenamefont{Irie,
  Watanabe, and Hashimoto}}]{irie2003nitrogen}
\bibinfo{author}{\bibfnamefont{H.}~\bibnamefont{Irie}},
  \bibinfo{author}{\bibfnamefont{Y.}~\bibnamefont{Watanabe}}, \bibnamefont{and}
  \bibinfo{author}{\bibfnamefont{K.}~\bibnamefont{Hashimoto}},
  \bibinfo{journal}{The Journal of Physical Chemistry B}
  \textbf{\bibinfo{volume}{107}}, \bibinfo{pages}{5483}
  (\bibinfo{year}{2003}{\natexlab{a}}).

\bibitem[{\citenamefont{Irie et~al.}(2003{\natexlab{b}})\citenamefont{Irie,
  Watanabe, and Hashimoto}}]{irie2003carbon}
\bibinfo{author}{\bibfnamefont{H.}~\bibnamefont{Irie}},
  \bibinfo{author}{\bibfnamefont{Y.}~\bibnamefont{Watanabe}}, \bibnamefont{and}
  \bibinfo{author}{\bibfnamefont{K.}~\bibnamefont{Hashimoto}},
  \bibinfo{journal}{Chemistry Letters} \textbf{\bibinfo{volume}{32}},
  \bibinfo{pages}{772} (\bibinfo{year}{2003}{\natexlab{b}}).

\bibitem[{\citenamefont{Li et~al.}(2005)\citenamefont{Li, Hwang, Lee, and
  Kim}}]{li2005synthesis}
\bibinfo{author}{\bibfnamefont{Y.}~\bibnamefont{Li}},
  \bibinfo{author}{\bibfnamefont{D.-S.} \bibnamefont{Hwang}},
  \bibinfo{author}{\bibfnamefont{N.~H.} \bibnamefont{Lee}}, \bibnamefont{and}
  \bibinfo{author}{\bibfnamefont{S.-J.} \bibnamefont{Kim}},
  \bibinfo{journal}{Chemical Physics Letters} \textbf{\bibinfo{volume}{404}},
  \bibinfo{pages}{25} (\bibinfo{year}{2005}).

\bibitem[{\citenamefont{Asahi et~al.}(2001)\citenamefont{Asahi, Morikawa,
  Ohwaki, Aoki, and Taga}}]{asahi2001visible}
\bibinfo{author}{\bibfnamefont{R.}~\bibnamefont{Asahi}},
  \bibinfo{author}{\bibfnamefont{T.}~\bibnamefont{Morikawa}},
  \bibinfo{author}{\bibfnamefont{T.}~\bibnamefont{Ohwaki}},
  \bibinfo{author}{\bibfnamefont{K.}~\bibnamefont{Aoki}}, \bibnamefont{and}
  \bibinfo{author}{\bibfnamefont{Y.}~\bibnamefont{Taga}},
  \bibinfo{journal}{science} \textbf{\bibinfo{volume}{293}},
  \bibinfo{pages}{269} (\bibinfo{year}{2001}).

\bibitem[{\citenamefont{Chen and Burda}(2004)}]{chen2004photoelectron}
\bibinfo{author}{\bibfnamefont{X.}~\bibnamefont{Chen}} \bibnamefont{and}
  \bibinfo{author}{\bibfnamefont{C.}~\bibnamefont{Burda}},
  \bibinfo{journal}{The Journal of Physical Chemistry B}
  \textbf{\bibinfo{volume}{108}}, \bibinfo{pages}{15446}
  (\bibinfo{year}{2004}).

\bibitem[{\citenamefont{Di~Valentin and Pacchioni}(2013)}]{catal_today_2013}
\bibinfo{author}{\bibfnamefont{C.}~\bibnamefont{Di~Valentin}} \bibnamefont{and}
  \bibinfo{author}{\bibfnamefont{G.}~\bibnamefont{Pacchioni}},
  \bibinfo{journal}{Catalysis today} \textbf{\bibinfo{volume}{206}},
  \bibinfo{pages}{12} (\bibinfo{year}{2013}).

\bibitem[{\citenamefont{Di~Valentin et~al.}(2005)\citenamefont{Di~Valentin,
  Pacchioni, and Selloni}}]{chem_mat_2005}
\bibinfo{author}{\bibfnamefont{C.}~\bibnamefont{Di~Valentin}},
  \bibinfo{author}{\bibfnamefont{G.}~\bibnamefont{Pacchioni}},
  \bibnamefont{and} \bibinfo{author}{\bibfnamefont{A.}~\bibnamefont{Selloni}},
  \bibinfo{journal}{Chemistry of Materials} \textbf{\bibinfo{volume}{17}},
  \bibinfo{pages}{6656} (\bibinfo{year}{2005}).

\bibitem[{\citenamefont{Ohno et~al.}(2003)\citenamefont{Ohno, Mitsui, and
  Matsumura}}]{ohno2003photocatalytic}
\bibinfo{author}{\bibfnamefont{T.}~\bibnamefont{Ohno}},
  \bibinfo{author}{\bibfnamefont{T.}~\bibnamefont{Mitsui}}, \bibnamefont{and}
  \bibinfo{author}{\bibfnamefont{M.}~\bibnamefont{Matsumura}},
  \bibinfo{journal}{Chemistry letters} \textbf{\bibinfo{volume}{32}},
  \bibinfo{pages}{364} (\bibinfo{year}{2003}).

\bibitem[{\citenamefont{Varley et~al.}(2011)\citenamefont{Varley, Janotti, and
  Van~de Walle}}]{varley2011mechanism}
\bibinfo{author}{\bibfnamefont{J.}~\bibnamefont{Varley}},
  \bibinfo{author}{\bibfnamefont{A.}~\bibnamefont{Janotti}}, \bibnamefont{and}
  \bibinfo{author}{\bibfnamefont{C.}~\bibnamefont{Van~de Walle}},
  \bibinfo{journal}{Advanced Materials} \textbf{\bibinfo{volume}{23}},
  \bibinfo{pages}{2343} (\bibinfo{year}{2011}).

\bibitem[{\citenamefont{Wang et~al.}(2009)\citenamefont{Wang, Wu, and
  Liu}}]{wang2009simple}
\bibinfo{author}{\bibfnamefont{H.}~\bibnamefont{Wang}},
  \bibinfo{author}{\bibfnamefont{Z.}~\bibnamefont{Wu}}, \bibnamefont{and}
  \bibinfo{author}{\bibfnamefont{Y.}~\bibnamefont{Liu}}, \bibinfo{journal}{The
  Journal of Physical Chemistry C} \textbf{\bibinfo{volume}{113}},
  \bibinfo{pages}{13317} (\bibinfo{year}{2009}).

\bibitem[{\citenamefont{Liu et~al.}(2010)\citenamefont{Liu, Sun, Smith, Wang,
  Lu, and Cheng}}]{liu2010sulfur}
\bibinfo{author}{\bibfnamefont{G.}~\bibnamefont{Liu}},
  \bibinfo{author}{\bibfnamefont{C.}~\bibnamefont{Sun}},
  \bibinfo{author}{\bibfnamefont{S.~C.} \bibnamefont{Smith}},
  \bibinfo{author}{\bibfnamefont{L.}~\bibnamefont{Wang}},
  \bibinfo{author}{\bibfnamefont{G.~Q.~M.} \bibnamefont{Lu}}, \bibnamefont{and}
  \bibinfo{author}{\bibfnamefont{H.-M.} \bibnamefont{Cheng}},
  \bibinfo{journal}{Journal of colloid and interface science}
  \textbf{\bibinfo{volume}{349}}, \bibinfo{pages}{477} (\bibinfo{year}{2010}).

\bibitem[{\citenamefont{Xie et~al.}(2018)\citenamefont{Xie, Li, and
  Xu}}]{xie2018enhanced}
\bibinfo{author}{\bibfnamefont{W.}~\bibnamefont{Xie}},
  \bibinfo{author}{\bibfnamefont{R.}~\bibnamefont{Li}}, \bibnamefont{and}
  \bibinfo{author}{\bibfnamefont{Q.}~\bibnamefont{Xu}},
  \bibinfo{journal}{Scientific reports} \textbf{\bibinfo{volume}{8}}
  (\bibinfo{year}{2018}).

\bibitem[{\citenamefont{Basera et~al.}(to be published)\citenamefont{Basera,
  Saini, Arora, Singh, Kumar, and Bhattacharya}}]{PB}
\bibinfo{author}{\bibfnamefont{P.}~\bibnamefont{Basera}},
  \bibinfo{author}{\bibfnamefont{S.}~\bibnamefont{Saini}},
  \bibinfo{author}{\bibfnamefont{E.}~\bibnamefont{Arora}},
  \bibinfo{author}{\bibfnamefont{A.}~\bibnamefont{Singh}},
  \bibinfo{author}{\bibfnamefont{M.}~\bibnamefont{Kumar}}, \bibnamefont{and}
  \bibinfo{author}{\bibfnamefont{S.}~\bibnamefont{Bhattacharya}}
  (\bibinfo{year}{to be published}).

\bibitem[{\citenamefont{Chiodo et~al.}(2010)\citenamefont{Chiodo,
  Garc{\'\i}a-Lastra, Iacomino, Ossicini, Zhao, Petek, and
  Rubio}}]{chiodo2010self}
\bibinfo{author}{\bibfnamefont{L.}~\bibnamefont{Chiodo}},
  \bibinfo{author}{\bibfnamefont{J.~M.} \bibnamefont{Garc{\'\i}a-Lastra}},
  \bibinfo{author}{\bibfnamefont{A.}~\bibnamefont{Iacomino}},
  \bibinfo{author}{\bibfnamefont{S.}~\bibnamefont{Ossicini}},
  \bibinfo{author}{\bibfnamefont{J.}~\bibnamefont{Zhao}},
  \bibinfo{author}{\bibfnamefont{H.}~\bibnamefont{Petek}}, \bibnamefont{and}
  \bibinfo{author}{\bibfnamefont{A.}~\bibnamefont{Rubio}},
  \bibinfo{journal}{Physical Review B} \textbf{\bibinfo{volume}{82}},
  \bibinfo{pages}{045207} (\bibinfo{year}{2010}).

\bibitem[{\citenamefont{Zhu and Gao}(2014)}]{zhu2014stability}
\bibinfo{author}{\bibfnamefont{T.}~\bibnamefont{Zhu}} \bibnamefont{and}
  \bibinfo{author}{\bibfnamefont{S.-P.} \bibnamefont{Gao}},
  \bibinfo{journal}{The Journal of Physical Chemistry C}
  \textbf{\bibinfo{volume}{118}}, \bibinfo{pages}{11385}
  (\bibinfo{year}{2014}).

\bibitem[{\citenamefont{Fuchs}(2005)}]{fuchs2005comparison}
\bibinfo{author}{\bibfnamefont{M.}~\bibnamefont{Fuchs}},
  \bibinfo{journal}{Density-Functional Theory Calculations for Modeling
  Materials and Bio-Molecular Properties and Functions-A Hands-On Computer
  Course, Los Angeles}  (\bibinfo{year}{2005}).

\bibitem[{\citenamefont{Perdew et~al.}(1996)\citenamefont{Perdew, Burke, and
  Ernzerhof}}]{perdew1996generalized}
\bibinfo{author}{\bibfnamefont{J.~P.} \bibnamefont{Perdew}},
  \bibinfo{author}{\bibfnamefont{K.}~\bibnamefont{Burke}}, \bibnamefont{and}
  \bibinfo{author}{\bibfnamefont{M.}~\bibnamefont{Ernzerhof}},
  \bibinfo{journal}{Physical review letters} \textbf{\bibinfo{volume}{77}},
  \bibinfo{pages}{3865} (\bibinfo{year}{1996}).

\bibitem[{\citenamefont{Hubbard}(1963)}]{hubbard1963electron}
\bibinfo{author}{\bibfnamefont{J.}~\bibnamefont{Hubbard}},
  \bibinfo{journal}{Proceedings of the Royal Society of London. Series A.
  Mathematical and Physical Sciences} \textbf{\bibinfo{volume}{276}},
  \bibinfo{pages}{238} (\bibinfo{year}{1963}).

\bibitem[{\citenamefont{Gerosa et~al.}(2017)\citenamefont{Gerosa, Bottani,
  Di~Valentin, Onida, and Pacchioni}}]{gerosa2017accuracy}
\bibinfo{author}{\bibfnamefont{M.}~\bibnamefont{Gerosa}},
  \bibinfo{author}{\bibfnamefont{C.}~\bibnamefont{Bottani}},
  \bibinfo{author}{\bibfnamefont{C.}~\bibnamefont{Di~Valentin}},
  \bibinfo{author}{\bibfnamefont{G.}~\bibnamefont{Onida}}, \bibnamefont{and}
  \bibinfo{author}{\bibfnamefont{G.}~\bibnamefont{Pacchioni}},
  \bibinfo{journal}{Journal of Physics: Condensed Matter}
  \textbf{\bibinfo{volume}{30}}, \bibinfo{pages}{044003}
  (\bibinfo{year}{2017}).

\bibitem[{\citenamefont{Kang and Hybertsen}(2010)}]{kang2010quasiparticle}
\bibinfo{author}{\bibfnamefont{W.}~\bibnamefont{Kang}} \bibnamefont{and}
  \bibinfo{author}{\bibfnamefont{M.~S.} \bibnamefont{Hybertsen}},
  \bibinfo{journal}{Physical Review B} \textbf{\bibinfo{volume}{82}},
  \bibinfo{pages}{085203} (\bibinfo{year}{2010}).

\bibitem[{\citenamefont{Jiang et~al.}(2010)\citenamefont{Jiang, Gomez-Abal,
  Rinke, and Scheffler}}]{jiang2010first}
\bibinfo{author}{\bibfnamefont{H.}~\bibnamefont{Jiang}},
  \bibinfo{author}{\bibfnamefont{R.~I.} \bibnamefont{Gomez-Abal}},
  \bibinfo{author}{\bibfnamefont{P.}~\bibnamefont{Rinke}}, \bibnamefont{and}
  \bibinfo{author}{\bibfnamefont{M.}~\bibnamefont{Scheffler}},
  \bibinfo{journal}{Physical Review B} \textbf{\bibinfo{volume}{82}},
  \bibinfo{pages}{045108} (\bibinfo{year}{2010}).

\bibitem[{\citenamefont{Karsai et~al.}(2014)\citenamefont{Karsai, Tiwald,
  Laskowski, Tran, Koller, Gr{\"a}fe, Burgd{\"o}rfer, Wirtz, and
  Blaha}}]{karsai2014f}
\bibinfo{author}{\bibfnamefont{F.}~\bibnamefont{Karsai}},
  \bibinfo{author}{\bibfnamefont{P.}~\bibnamefont{Tiwald}},
  \bibinfo{author}{\bibfnamefont{R.}~\bibnamefont{Laskowski}},
  \bibinfo{author}{\bibfnamefont{F.}~\bibnamefont{Tran}},
  \bibinfo{author}{\bibfnamefont{D.}~\bibnamefont{Koller}},
  \bibinfo{author}{\bibfnamefont{S.}~\bibnamefont{Gr{\"a}fe}},
  \bibinfo{author}{\bibfnamefont{J.}~\bibnamefont{Burgd{\"o}rfer}},
  \bibinfo{author}{\bibfnamefont{L.}~\bibnamefont{Wirtz}}, \bibnamefont{and}
  \bibinfo{author}{\bibfnamefont{P.}~\bibnamefont{Blaha}},
  \bibinfo{journal}{Physical Review B} \textbf{\bibinfo{volume}{89}},
  \bibinfo{pages}{125429} (\bibinfo{year}{2014}).

\bibitem[{\citenamefont{Onida et~al.}(2002)\citenamefont{Onida, Reining, and
  Rubio}}]{onida2002electronic}
\bibinfo{author}{\bibfnamefont{G.}~\bibnamefont{Onida}},
  \bibinfo{author}{\bibfnamefont{L.}~\bibnamefont{Reining}}, \bibnamefont{and}
  \bibinfo{author}{\bibfnamefont{A.}~\bibnamefont{Rubio}},
  \bibinfo{journal}{Reviews of Modern Physics} \textbf{\bibinfo{volume}{74}},
  \bibinfo{pages}{601} (\bibinfo{year}{2002}).

\bibitem[{\citenamefont{Fuchs et~al.}(2008)\citenamefont{Fuchs, R{\"o}dl,
  Schleife, and Bechstedt}}]{fuchs2008efficient}
\bibinfo{author}{\bibfnamefont{F.}~\bibnamefont{Fuchs}},
  \bibinfo{author}{\bibfnamefont{C.}~\bibnamefont{R{\"o}dl}},
  \bibinfo{author}{\bibfnamefont{A.}~\bibnamefont{Schleife}}, \bibnamefont{and}
  \bibinfo{author}{\bibfnamefont{F.}~\bibnamefont{Bechstedt}},
  \bibinfo{journal}{Physical Review B} \textbf{\bibinfo{volume}{78}},
  \bibinfo{pages}{085103} (\bibinfo{year}{2008}).

\bibitem[{\citenamefont{Salpeter and Bethe}(1951)}]{salpeter1951relativistic}
\bibinfo{author}{\bibfnamefont{E.~E.} \bibnamefont{Salpeter}} \bibnamefont{and}
  \bibinfo{author}{\bibfnamefont{H.~A.} \bibnamefont{Bethe}},
  \bibinfo{journal}{Physical Review} \textbf{\bibinfo{volume}{84}},
  \bibinfo{pages}{1232} (\bibinfo{year}{1951}).

\bibitem[{\citenamefont{Bl{\"o}chl}(1994)}]{blochl1994projector}
\bibinfo{author}{\bibfnamefont{P.~E.} \bibnamefont{Bl{\"o}chl}},
  \bibinfo{journal}{Physical review B} \textbf{\bibinfo{volume}{50}},
  \bibinfo{pages}{17953} (\bibinfo{year}{1994}).

\bibitem[{\citenamefont{Kresse and
  Furthm{\"u}ller}(1996)}]{kresse1996efficient}
\bibinfo{author}{\bibfnamefont{G.}~\bibnamefont{Kresse}} \bibnamefont{and}
  \bibinfo{author}{\bibfnamefont{J.}~\bibnamefont{Furthm{\"u}ller}},
  \bibinfo{journal}{Physical review B} \textbf{\bibinfo{volume}{54}},
  \bibinfo{pages}{11169} (\bibinfo{year}{1996}).

\bibitem[{\citenamefont{Monkhorst and Pack}(1976)}]{monkhorst1976special}
\bibinfo{author}{\bibfnamefont{H.~J.} \bibnamefont{Monkhorst}}
  \bibnamefont{and} \bibinfo{author}{\bibfnamefont{J.~D.} \bibnamefont{Pack}},
  \bibinfo{journal}{Physical review B} \textbf{\bibinfo{volume}{13}},
  \bibinfo{pages}{5188} (\bibinfo{year}{1976}).

\bibitem[{\citenamefont{Hedin}(1965)}]{hedin1965new}
\bibinfo{author}{\bibfnamefont{L.}~\bibnamefont{Hedin}},
  \bibinfo{journal}{Physical Review} \textbf{\bibinfo{volume}{139}},
  \bibinfo{pages}{A796} (\bibinfo{year}{1965}).

\bibitem[{\citenamefont{Hybertsen}(1985)}]{hybertsen1985ms}
\bibinfo{author}{\bibfnamefont{M.}~\bibnamefont{Hybertsen}},
  \bibinfo{journal}{Phys. Rev. Lett.} \textbf{\bibinfo{volume}{55}},
  \bibinfo{pages}{1418} (\bibinfo{year}{1985}).

\bibitem[{\citenamefont{Wemple}(1977)}]{wemple1977optical}
\bibinfo{author}{\bibfnamefont{S.}~\bibnamefont{Wemple}}, \bibinfo{journal}{The
  Journal of Chemical Physics} \textbf{\bibinfo{volume}{67}},
  \bibinfo{pages}{2151} (\bibinfo{year}{1977}).

\bibitem[{\citenamefont{Lyons et~al.}(2014)\citenamefont{Lyons, Janotti, and
  Van~de Walle}}]{lyons2014effects}
\bibinfo{author}{\bibfnamefont{J.}~\bibnamefont{Lyons}},
  \bibinfo{author}{\bibfnamefont{A.}~\bibnamefont{Janotti}}, \bibnamefont{and}
  \bibinfo{author}{\bibfnamefont{C.}~\bibnamefont{Van~de Walle}},
  \bibinfo{journal}{Physical Review B} \textbf{\bibinfo{volume}{89}},
  \bibinfo{pages}{035204} (\bibinfo{year}{2014}).

\bibitem[{\citenamefont{Rinke et~al.}(2012)\citenamefont{Rinke, Schleife,
  Kioupakis, Janotti, R{\"o}dl, Bechstedt, Scheffler, and Van~de
  Walle}}]{rinke2012first}
\bibinfo{author}{\bibfnamefont{P.}~\bibnamefont{Rinke}},
  \bibinfo{author}{\bibfnamefont{A.}~\bibnamefont{Schleife}},
  \bibinfo{author}{\bibfnamefont{E.}~\bibnamefont{Kioupakis}},
  \bibinfo{author}{\bibfnamefont{A.}~\bibnamefont{Janotti}},
  \bibinfo{author}{\bibfnamefont{C.}~\bibnamefont{R{\"o}dl}},
  \bibinfo{author}{\bibfnamefont{F.}~\bibnamefont{Bechstedt}},
  \bibinfo{author}{\bibfnamefont{M.}~\bibnamefont{Scheffler}},
  \bibnamefont{and} \bibinfo{author}{\bibfnamefont{C.~G.} \bibnamefont{Van~de
  Walle}}, \bibinfo{journal}{Physical review letters}
  \textbf{\bibinfo{volume}{108}}, \bibinfo{pages}{126404}
  (\bibinfo{year}{2012}).

\bibitem[{\citenamefont{Alkauskas et~al.}(2016)\citenamefont{Alkauskas,
  McCluskey, and Van~de Walle}}]{alkauskas2016tutorial}
\bibinfo{author}{\bibfnamefont{A.}~\bibnamefont{Alkauskas}},
  \bibinfo{author}{\bibfnamefont{M.~D.} \bibnamefont{McCluskey}},
  \bibnamefont{and} \bibinfo{author}{\bibfnamefont{C.~G.} \bibnamefont{Van~de
  Walle}}, \bibinfo{journal}{Journal of Applied Physics}
  \textbf{\bibinfo{volume}{119}}, \bibinfo{pages}{181101}
  (\bibinfo{year}{2016}).

\bibitem[{\citenamefont{Li et~al.}(2007)\citenamefont{Li, Zhang, Huo, and
  Zhu}}]{li2007supercritical}
\bibinfo{author}{\bibfnamefont{H.}~\bibnamefont{Li}},
  \bibinfo{author}{\bibfnamefont{X.}~\bibnamefont{Zhang}},
  \bibinfo{author}{\bibfnamefont{Y.}~\bibnamefont{Huo}}, \bibnamefont{and}
  \bibinfo{author}{\bibfnamefont{J.}~\bibnamefont{Zhu}},
  \bibinfo{journal}{Environmental science \& technology}
  \textbf{\bibinfo{volume}{41}}, \bibinfo{pages}{4410} (\bibinfo{year}{2007}).

\bibitem[{\citenamefont{Cheung et~al.}(2008)\citenamefont{Cheung, Nachimuthu,
  Engelhard, Wang, and Chambers}}]{cheung2008n}
\bibinfo{author}{\bibfnamefont{S.~H.} \bibnamefont{Cheung}},
  \bibinfo{author}{\bibfnamefont{P.}~\bibnamefont{Nachimuthu}},
  \bibinfo{author}{\bibfnamefont{M.~H.} \bibnamefont{Engelhard}},
  \bibinfo{author}{\bibfnamefont{C.~M.} \bibnamefont{Wang}}, \bibnamefont{and}
  \bibinfo{author}{\bibfnamefont{S.~A.} \bibnamefont{Chambers}},
  \bibinfo{journal}{Surface Science} \textbf{\bibinfo{volume}{602}},
  \bibinfo{pages}{133} (\bibinfo{year}{2008}).

\bibitem[{\citenamefont{Lin et~al.}(2011)\citenamefont{Lin, Rong, Ji, and
  Fu}}]{lin2011carbon}
\bibinfo{author}{\bibfnamefont{X.}~\bibnamefont{Lin}},
  \bibinfo{author}{\bibfnamefont{F.}~\bibnamefont{Rong}},
  \bibinfo{author}{\bibfnamefont{X.}~\bibnamefont{Ji}}, \bibnamefont{and}
  \bibinfo{author}{\bibfnamefont{D.}~\bibnamefont{Fu}},
  \bibinfo{journal}{Microporous and Mesoporous Materials}
  \textbf{\bibinfo{volume}{142}}, \bibinfo{pages}{276} (\bibinfo{year}{2011}).

\bibitem[{\citenamefont{Gali et~al.}(2012)\citenamefont{Gali, Coulter, and
  Manousakis}}]{gali2012ab}
\bibinfo{author}{\bibfnamefont{A.}~\bibnamefont{Gali}},
  \bibinfo{author}{\bibfnamefont{J.~E.} \bibnamefont{Coulter}},
  \bibnamefont{and}
  \bibinfo{author}{\bibfnamefont{E.}~\bibnamefont{Manousakis}},
  \bibinfo{journal}{arXiv preprint arXiv:1207.6098}  (\bibinfo{year}{2012}).

\bibitem[{\citenamefont{Heyd et~al.}(2003)\citenamefont{Heyd, Scuseria, and
  Ernzerhof}}]{heyd2003hybrid}
\bibinfo{author}{\bibfnamefont{J.}~\bibnamefont{Heyd}},
  \bibinfo{author}{\bibfnamefont{G.~E.} \bibnamefont{Scuseria}},
  \bibnamefont{and}
  \bibinfo{author}{\bibfnamefont{M.}~\bibnamefont{Ernzerhof}},
  \bibinfo{journal}{The Journal of Chemical Physics}
  \textbf{\bibinfo{volume}{118}}, \bibinfo{pages}{8207} (\bibinfo{year}{2003}).

\bibitem[{\citenamefont{Krukau et~al.}(2008)\citenamefont{Krukau, Scuseria,
  Perdew, and Savin}}]{krukau2008hybrid}
\bibinfo{author}{\bibfnamefont{A.~V.} \bibnamefont{Krukau}},
  \bibinfo{author}{\bibfnamefont{G.~E.} \bibnamefont{Scuseria}},
  \bibinfo{author}{\bibfnamefont{J.~P.} \bibnamefont{Perdew}},
  \bibnamefont{and} \bibinfo{author}{\bibfnamefont{A.}~\bibnamefont{Savin}},
  \bibinfo{journal}{The Journal of chemical physics}
  \textbf{\bibinfo{volume}{129}}, \bibinfo{pages}{124103}
  (\bibinfo{year}{2008}).

\bibitem[{\citenamefont{De{\'a}k et~al.}(2010)\citenamefont{De{\'a}k, Aradi,
  Frauenheim, Janz{\'e}n, and Gali}}]{deak2010accurate}
\bibinfo{author}{\bibfnamefont{P.}~\bibnamefont{De{\'a}k}},
  \bibinfo{author}{\bibfnamefont{B.}~\bibnamefont{Aradi}},
  \bibinfo{author}{\bibfnamefont{T.}~\bibnamefont{Frauenheim}},
  \bibinfo{author}{\bibfnamefont{E.}~\bibnamefont{Janz{\'e}n}},
  \bibnamefont{and} \bibinfo{author}{\bibfnamefont{A.}~\bibnamefont{Gali}},
  \bibinfo{journal}{Physical Review B} \textbf{\bibinfo{volume}{81}},
  \bibinfo{pages}{153203} (\bibinfo{year}{2010}).

\bibitem[{\citenamefont{Gonzalez et~al.}(1997)\citenamefont{Gonzalez, Zallen,
  and Berger}}]{gonzalez1997infrared}
\bibinfo{author}{\bibfnamefont{R.}~\bibnamefont{Gonzalez}},
  \bibinfo{author}{\bibfnamefont{R.}~\bibnamefont{Zallen}}, \bibnamefont{and}
  \bibinfo{author}{\bibfnamefont{H.}~\bibnamefont{Berger}},
  \bibinfo{journal}{Physical Review B} \textbf{\bibinfo{volume}{55}},
  \bibinfo{pages}{7014} (\bibinfo{year}{1997}).

\bibitem[{\citenamefont{Hosaka et~al.}(1997)\citenamefont{Hosaka, Sekiya,
  Satoko, and Kurita}}]{hosaka1997optical}
\bibinfo{author}{\bibfnamefont{N.}~\bibnamefont{Hosaka}},
  \bibinfo{author}{\bibfnamefont{T.}~\bibnamefont{Sekiya}},
  \bibinfo{author}{\bibfnamefont{C.}~\bibnamefont{Satoko}}, \bibnamefont{and}
  \bibinfo{author}{\bibfnamefont{S.}~\bibnamefont{Kurita}},
  \bibinfo{journal}{Journal of the Physical Society of Japan}
  \textbf{\bibinfo{volume}{66}}, \bibinfo{pages}{877} (\bibinfo{year}{1997}).

\bibitem[{\citenamefont{Atambo et~al.}(2019)\citenamefont{Atambo, Varsano,
  Ferretti, Ataei, Caldas, Molinari, and Selloni}}]{atambo2019electronic}
\bibinfo{author}{\bibfnamefont{M.~O.} \bibnamefont{Atambo}},
  \bibinfo{author}{\bibfnamefont{D.}~\bibnamefont{Varsano}},
  \bibinfo{author}{\bibfnamefont{A.}~\bibnamefont{Ferretti}},
  \bibinfo{author}{\bibfnamefont{S.~S.} \bibnamefont{Ataei}},
  \bibinfo{author}{\bibfnamefont{M.~J.} \bibnamefont{Caldas}},
  \bibinfo{author}{\bibfnamefont{E.}~\bibnamefont{Molinari}}, \bibnamefont{and}
  \bibinfo{author}{\bibfnamefont{A.}~\bibnamefont{Selloni}},
  \bibinfo{journal}{Physical Review Materials} \textbf{\bibinfo{volume}{3}},
  \bibinfo{pages}{045401} (\bibinfo{year}{2019}).

%




\end{thebibliography}

\end{document}